\documentclass[twocolumn,secnumarabic,amssymb,nobibnotes,aps,pre]{revtex4-1}

\usepackage{array}
\usepackage{graphicx}		 
\usepackage{amsmath}		
\usepackage[colorlinks=false,linkcolor=blue]{hyperref}
\begin{document}
\title{Strange nonchaotic attractors for computation}
\author{M. Sathish Aravindh$^{1,2}$, A. Venkatesan$^{1}$, M. Lakshmanan$^{2}$} %
\affiliation{$^{1}$PG \& Research Department of Physics, Nehru Memorial College (Autonomous), Puthanampatti, Tiruchirappalli - 621 007, India.\\ $^{2}$Centre for Nonlinear Dynamics, School of Physics, Bharathidasan University, Tiruchirappalli - 620 024, India.}
\email{sathisharavindhm@gmail.com, av.phys@gmail.com,\\ lakshman.cnld@gmail.com} 

\begin{abstract}
We investigate the response of quasiperiodically driven nonlinear systems exhibiting strange nonchaotic attractors (SNAs) to deterministic input signals. We show that if one uses two square waves in aperiodic manner as input to a quasiperiodically driven double-well Duffing oscillator system, the response of the system can produce logical output controlled by such a forcing. Changing the threshold or biasing of the system changes the output to another logic operation and memory latch. The interplay of nonlinearity and quasiperiodic forcing yields logic behaviour and the emergent outcome of such a system is a logic gate. It is further shown that the logical behaviours persist even for experimental noise floor. Thus the SNA turns out to be an efficient tool for computation.
\end{abstract}
\maketitle
\section{Introduction}
Strange nonchaotic attractors (SNAs) are attractors which possess fractal geometry but exhibit no sensitive dependence on initial conditions. SNAs occur in all dissipative dynamical systems when the attractors formed at the accumulation points of period-doubling cascades are fractal sets with zero Lyapunov exponent. Such attractors are however not physically observable because the set of parameter values for them to arise has Lebesgue measure zero in the parameter space. Situations where SNAs can arise typically were described by Grebogi \textit{et al.} \cite{grebogi1984strange}, who found that quasiperiodically driven dynamical systems admit SNAs in parameter regions of positive Lebesgue measure. Since then, there have been many studies on SNAs in quasiperiodically forced systems \cite{ditto1990experimental,zhou1992observation, *thamilmaran2006experimental,ding1997observation, ruiz2007experimental, bezruchko2000experimental, bondeson1985quasiperiodically, ding1994phase,*prasad2003strange,
zhou1997robust, *ramaswamy1997synchronization, lindner2015strange, *lindner2016simple, heagy1994birth, *nishikawa1996fractalization, *yalccinkaya1996blowout, *lai1996transition, *prasad1997prasad, *witt1997birth, *prasad1999collision, *wang2004strange, venkatesan2000intermittency, prasad2001strange,
venkatesan1999birth, *venkatesan2001interruption, *gopal2013applicability}. Experimental observations of SNAs have been reported in a quasiperiodically driven magnetoelastic ribbon system \cite{ditto1990experimental}, in electronic circuits \cite{zhou1992observation, *thamilmaran2006experimental}, in a plasma system \cite{ding1997observation}, in an electrochemical cell \cite{ruiz2007experimental} and in a system near the torus-doubling critical point \cite{bezruchko2000experimental}. 

Physically, SNAs are relevant to situations such as localization of quantum particles in spatially quasiperiodic potential systems \cite{bondeson1985quasiperiodically}. These exotic attractors are also important for biological systems \cite{ding1994phase, *prasad2003strange} and they may be useful for nonlinear dynamics based communication as well \cite{zhou1997robust,*ramaswamy1997synchronization}. Recently evidence for strange nonchaotic behaviour has been identified in  the pulsation of stars like KIC 5520878 \cite{lindner2015strange, *lindner2016simple}. Most of the works in the literature considered the process by which an SNA can be created from a regular attractor, or its disappearance in the transition to a chaotic attractor \cite{heagy1994birth, *nishikawa1996fractalization, *yalccinkaya1996blowout, *lai1996transition, *prasad1997prasad, *witt1997birth, *prasad1999collision, *wang2004strange, venkatesan2000intermittency, 
prasad2001strange, venkatesan1999birth, *venkatesan2001interruption, *gopal2013applicability}. SNAs can also be quantitatively characterized by a variety of measures/methods including Lyapunov exponents, fractal dimension and spectral properties as well as examination of time series \cite{lindner2015strange, *lindner2016simple, heagy1994birth, *nishikawa1996fractalization, *yalccinkaya1996blowout, *lai1996transition, *prasad1997prasad, *witt1997birth, *prasad1999collision, *wang2004strange, 
venkatesan2000intermittency, prasad2001strange, venkatesan1999birth, *venkatesan2001interruption, *gopal2013applicability, pikovsky1995singular, *pikovsky1995characterizing}. The geometric strangeness of the attractor can be measured through indices such as the phase sensitivity exponent, while the chaoticity property can be checked by examining the finite-time Lyapunov exponents  \cite{pikovsky1995singular, *pikovsky1995characterizing}. Mathematical issues concerning the generation and properties of SNAs have also been addressed  \cite{stark1997invariant, *sturman2000semi}. In this regard, if an SNA can persist under small perturbations, it is said to be robust \cite{hunt2001fractal, *kim2003fractal}. So far robust SNAs have been identified and studied extensively in quasiperiodically driven dynamical systems \cite{stark1997invariant, *sturman2000semi, hunt2001fractal, *kim2003fractal}.

It is well known that the approaching of physical limits on Moore's law has led to the development of alternative methods to perform more number of computations out of limited number of hardwares \cite{sinha1998dynamics, *sinha1999computing, prusha1999nonlinearity, murali2003implementation, ashwin2004encoding, *ashwin2005discrete, *bick2009dynamical, *neves2012computation,*neves2017noise,  murali2009reliable, munakata2002chaos,*sinha2002flexible, murali2007using, 	venkatesh2016analytical, gupta2011noise, kohar2012noise, *kohar2014enhanced}.  In this direction, in 1998, the important work of Sinha and Ditto paved a new avenue of using chaos for computation \cite{sinha1998dynamics, *sinha1999computing}. They have proposed a chaos-computing scheme based on the thresholding method to achieve controlled response from a chaotic system. Simultaneously, Prusha and Lindner emphasized the importance of nonlinearity over chaos using a nonlinear paramaterized map and illustrated why chaos and computation require nonlinearity \cite{prusha1999nonlinearity}. Munakata \textit{et al.} realized various logic operations by employing a single chaotic element, especially a 1-D chaotic dynamical system (logistic map) \cite{munakata2002chaos,*sinha2002flexible}. Murali \textit{et al.} reported experimental realization of a fundamental NOR gate using chaotic systems \cite{murali2003implementation}. 

Further Murali \textit{et al.} proposed different schemes to obtain key logic structures by using synchronization \cite{murali2007using,	venkatesh2016analytical} and stochastic  resonance \cite{murali2009reliable} of nonlinear systems. Following this, Kohar \textit{et al.} have enhanced our understanding of nonlinear computing by adding either additional positive or negative asymmetric biasing to a bistable system driven by two input signals which yield logic function of two signals in an optimal window of moderate noise \cite{kohar2012noise,*kohar2014enhanced}. The phenomenon of Logical Stochastic Resonance (LSR) has been realized theoretically and experimentally in diverse systems, namely a nanoscale device \cite{guerra2010noise}, resonant tunnel diodes \cite{worschech2010universal}, a vertical cavity surface emitting laser \cite{zamora2010numerical, *zhang2010effect}, a polarization bistable laser \cite{singh2011enhancement}, a chemical system \cite{bulsara2010logical}, synthetic gene networks \cite{dari2011creating}, and so on. Recently, two of the present authors  along with Venkatesh employed coupled dynamical systems to build dynamical logic gates by altering the value of the logic inputs \cite{venkatesh2017design, *venkatesh2017implementation}.

From a different point of view, Kia \textit{et al.} demonstrated how unstable periodic orbits can be exploited to model chaos computing \cite{kia2011unstable}. Later on
Kia \textit{et al.} have also shown how the inherent noise reduction properties in coupled systems can be used for computation \cite{kia2014noise}. Further Wang and Roychowdhury showed that self sustaining oscillators of any type can function as latches and registers if Boolean logic states are represented as the phase of oscillatory signals \cite{wang2014phlogon, *wang2015design, *roychowdhury2015boolean, *wang2017rigorous}. Borresen and Lynch have demonstrated how coupled threshold oscillators may be used as the principle components of computers \cite{borresen2012oscillatory}. It has also been pointed out that heteroclinic computing offers a paradigm for computation by collective system of nonlinear oscillators \cite{ashwin2004encoding, *ashwin2005discrete, *bick2009dynamical, *neves2012computation,*neves2017noise}. 

These approaches have been used to implement different  types of logic operations in a single set of nonlinear systems rather than needing multiple  hardwares for different types of computations \cite{sinha1998dynamics, *sinha1999computing, prusha1999nonlinearity, murali2003implementation, ashwin2004encoding, *ashwin2005discrete, *bick2009dynamical, *neves2012computation,*neves2017noise, murali2009reliable, venkatesh2017design, *venkatesh2017implementation}. Although, a nonlinear dynamical system can be a processor of flexibly configured and reconfigured device to emulate different logic gates, it was shown that the manufacturing non-idealities and ambient noise make it difficult to obtain different logic  functions in these systems \cite{dari2011logical, *kia2015coupling, *kia2017nonlinear, *kohar2017implementing}. In fact chaotic systems are highly sensitive to initial perturbations and thus a small amount of noise can completely change the system dynamics. As a result, special attention needs to be paid in choosing the appropriate nonlinear dynamics based computing systems which are robust against noise \cite{kia2011unstable}.

In the present paper, we propose a new and simple approach to encapsulate computations and noise robustness at the dynamics level. In particular, we present a route to logical SNAs in quasiperiodically driven nonlinear oscillator systems. We show that if we use two square waves in an aperiodic manner as input to a quasiperiodically driven double-well Duffing oscillator, the response of the oscillator can produce logical output controlled by such a forcing. Changing the threshold or biasing of the system changes the output to another logic operation and memory latch. We also show how by using such robust SNAs, including even noise, one can emulate different logic functions and thereby providing a sound nonlinear dynamics basis for computation.
  
The plan of the paper is as follows. In sec. II we analyse the dynamics of quasiperiodically forced double-well Duffing oscillator. In sec. III we discuss the effect of three level square waves on quasiperiodically driven Duffing oscillator, the mechanism of logical SNA and the characterization of logical SNA. In sec. IV we discuss implementation of other logic gates and SR flip flop. We also analyse the effect of noise on the logic gates. Finally in sec.V we present our conclusion.

\section{Dynamics of quasiperiodically forced Double-well Duffing oscillator}

To illustrate our findings, we consider the quasiperiodically driven Duffing oscillator,
\begin{align}
\dot{x}&=y  \nonumber \\
\dot{y}&=-\alpha \dot{x} - \beta (x^3-x)+ A(\sin \theta +\sin \phi)+I+\varepsilon +\sqrt{D}\xi(t)  \nonumber \\
\dot{\theta}&= \omega_1, ~~\dot{\phi}= \omega_2 
\label{equ1}
\end{align}                                                       
The simplest experimental realization of (\ref{equ1}), namely magnetoelastic ribbon \cite{ditto1990experimental}, has been studied extensively for SNA \cite{venkatesan2000intermittency, lakshmanan2003chaos}. The quantities $A$, $\omega_1$ and $\omega_2$ in (\ref{equ1}) correspond to the  amplitude and frequencies of the external two-frequency forcing. $\varepsilon$ is  an asymmetric bias input. $I$ is the low amplitude input square wave signal. $\xi(t)$ is a Gaussian white noise of intensity $D$. 

\begin{figure}[h]
	\centering
	\includegraphics[width=0.48\linewidth]{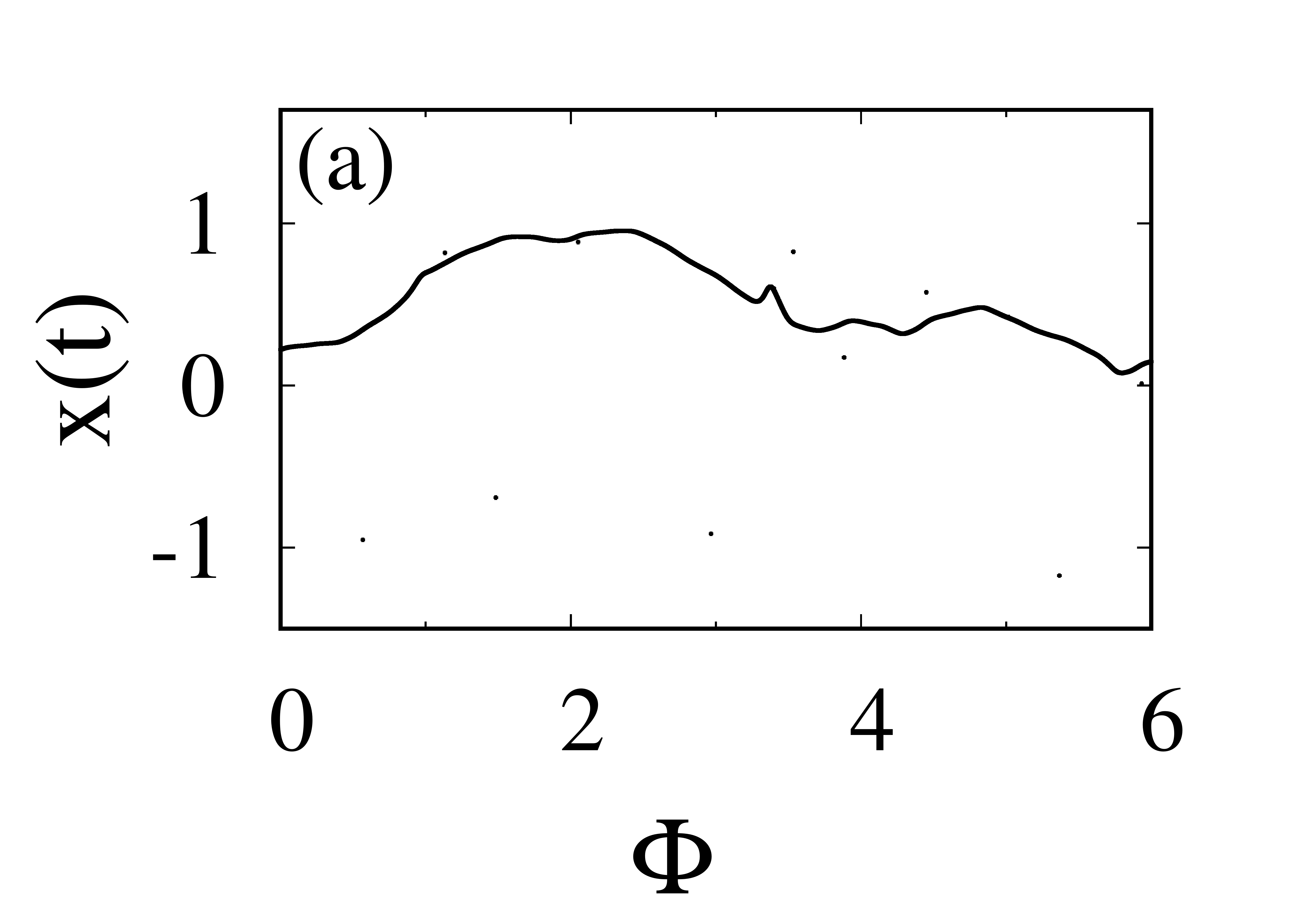}
	\includegraphics[width=0.48\linewidth]{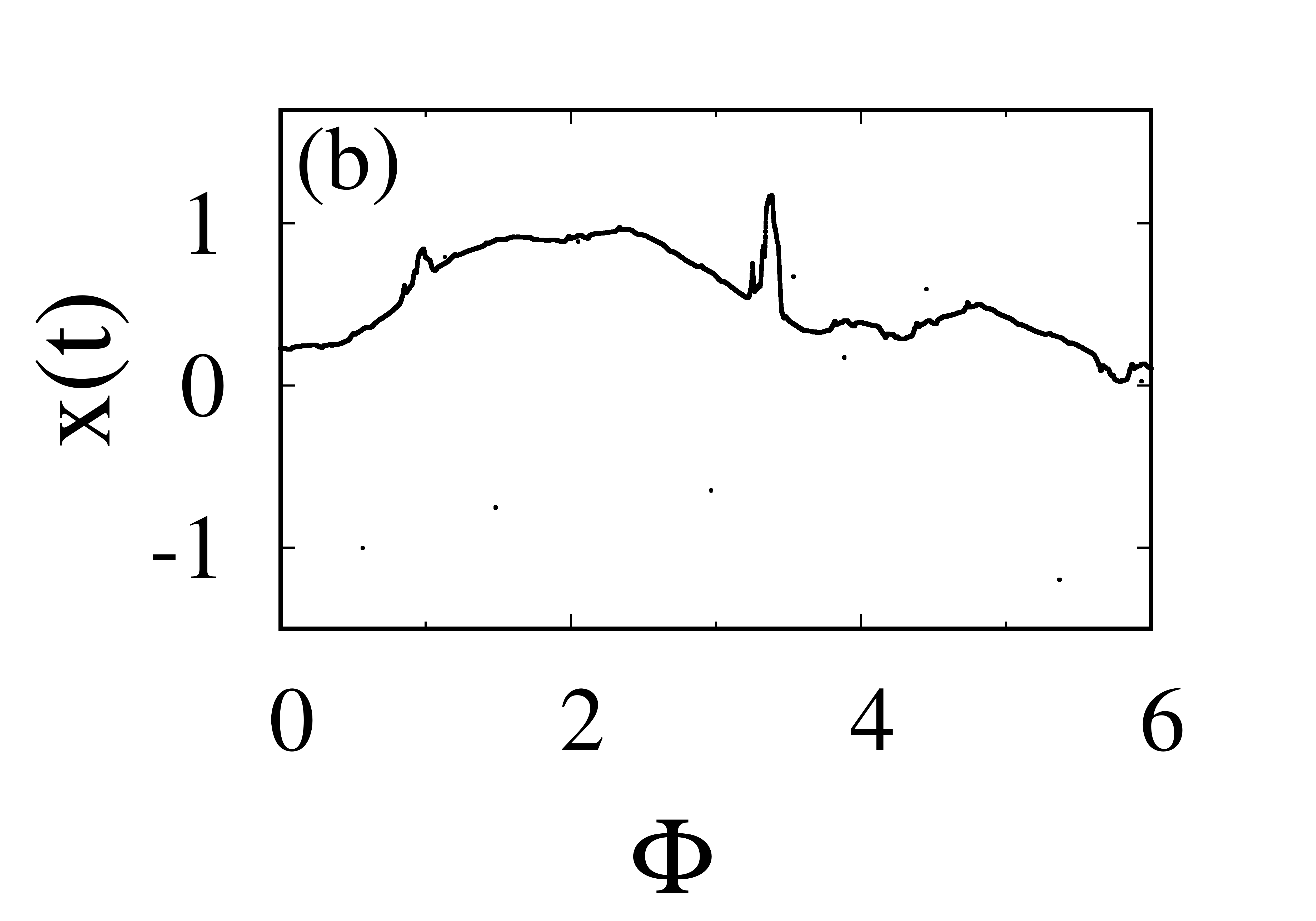}
	\includegraphics[width=0.48\linewidth]{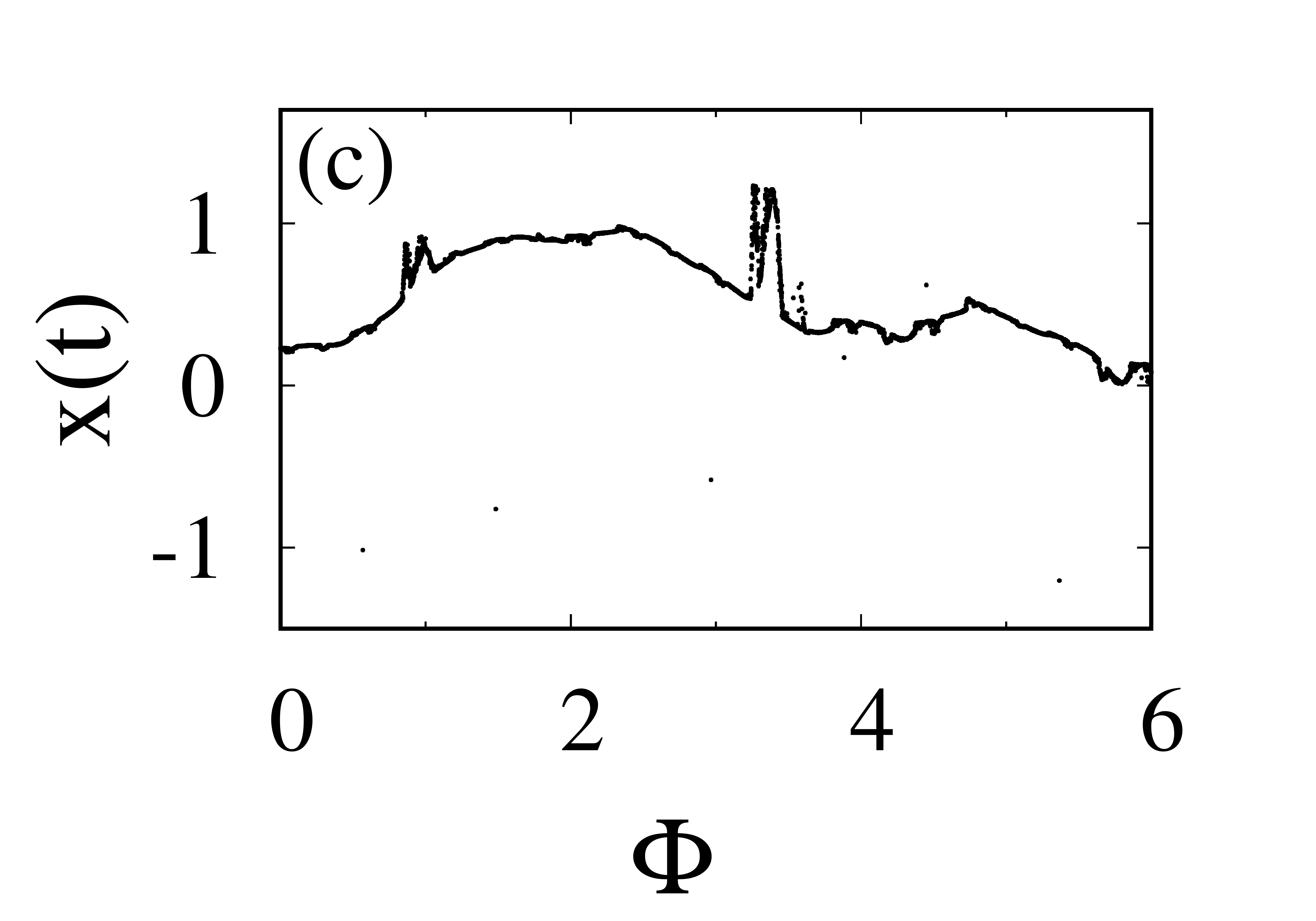} 
	\includegraphics[width=0.48\linewidth]{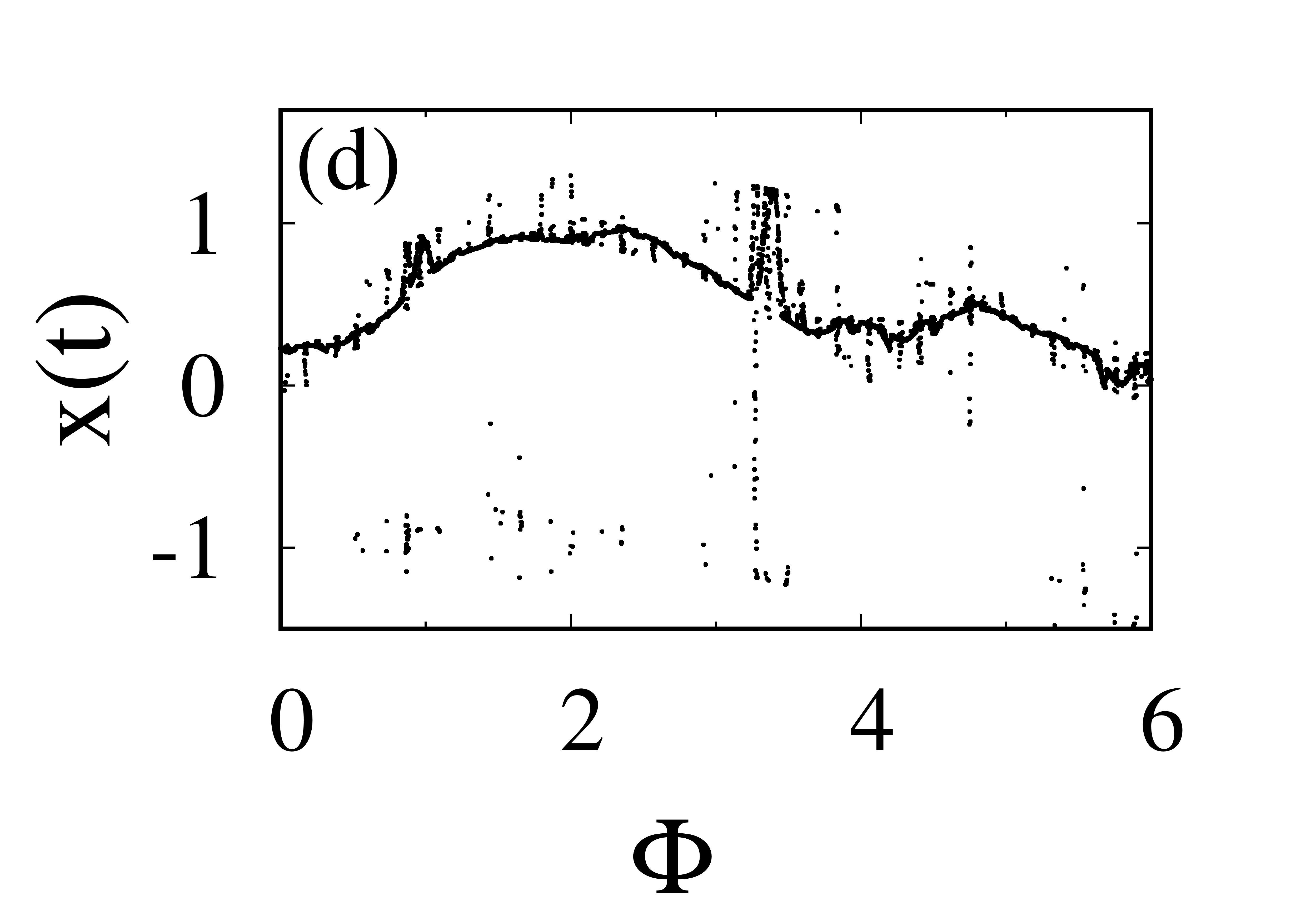}
	\caption{Projection of the numerically simulated attractors of Eq.\eqref{equ1} in the ($\phi$ -x) plane for various values of 'A' (a) torus for A=0.3110 (b) wrinkled torus for A= 0.31120 (c)  SNA for A= 0.311227 (d) chaos  for A=0.31124 and $\varepsilon$=0.05, when $I=0$ and $D=0.0$}
	\label{fig1}
\end{figure}

The existence of logical behaviour observed in the present study of Eq.\eqref{equ1} suggests that there may be experimental realization of different types of logic functions which deserve further study. For this purpose, we first study the dynamics of \eqref{equ1} in the absence of noise and input square waves. For our numerical calculation, we fix the parameters as $\alpha=0.5,~ \beta=1.0,~\omega_1=1.0,~\omega_2 =\frac{1}{2}(\sqrt{5}-1)$, $\epsilon=0.05$ and vary $'A'$. To visualize the attractor, it is convenient to use Poincar\'e surface of section technique. Specifically, we sample the system at time intervals corresponding to the variable $\theta_{n}= \omega_{1}t_{n}=2\pi n$, where $n=0,1,2...$. We then examine the dynamical variables $\phi_{n}(\text{mod} 2\pi)$ and $x_{n}$ on the surface of section. This is shown in Fig.\ref{fig1} which clearly portrays the transition from torus to SNA as a function of A. For A=0.311, the attractor is a  torus and can be seen as a single smooth strand in the Poincar\'e surface of section plot in its $(\phi-x)$ plane (Fig.\ref{fig1}(a)). As A is increased further to A=0.3112, one obtains the strand shown in Fig.\ref{fig1}(b), which loses its smoothness and becomes a wrinkled attractor. On increasing A to A=0.311227, the nature of the attractor becomes fractal and a SNA as shown in Fig.\ref{fig1}(c). Finally for A=0.31124, the attractor becomes a chaotic one (see Fig.\ref{fig1}(d)).

\begin{figure}
	\centering
	\includegraphics[width=0.95\linewidth]{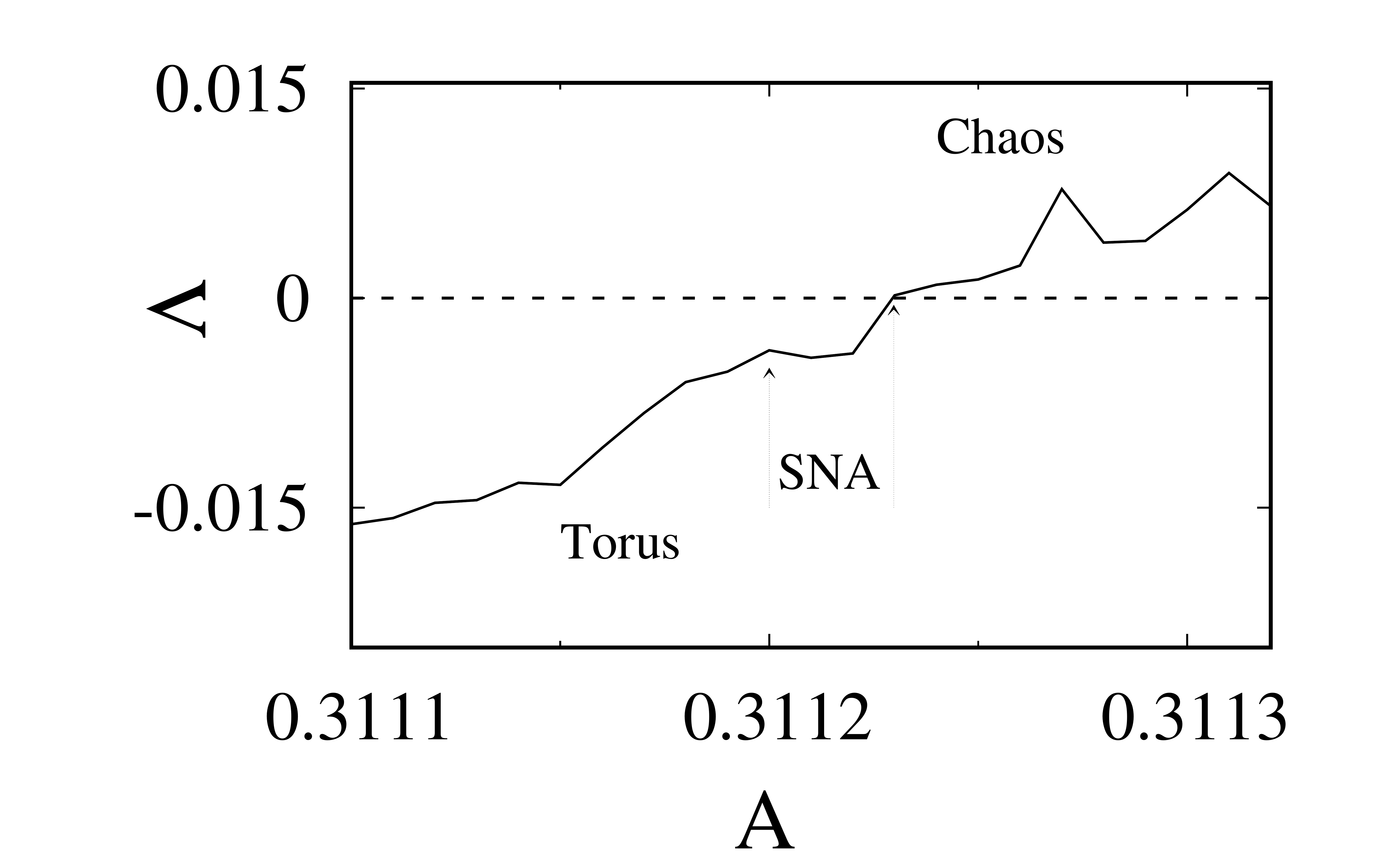}
	\caption{Maximal Lyapunov exponent  '$\Lambda$' versus  control parameter 'A', in the absence of inputs $I_{1}$ and $I_{2}$ with bias $\varepsilon=0.05$ and with no noise (D=0.0).}
	\label{fig2}
\end{figure}

\begin{figure}
	\centering
	
	\includegraphics[width=0.9\linewidth]{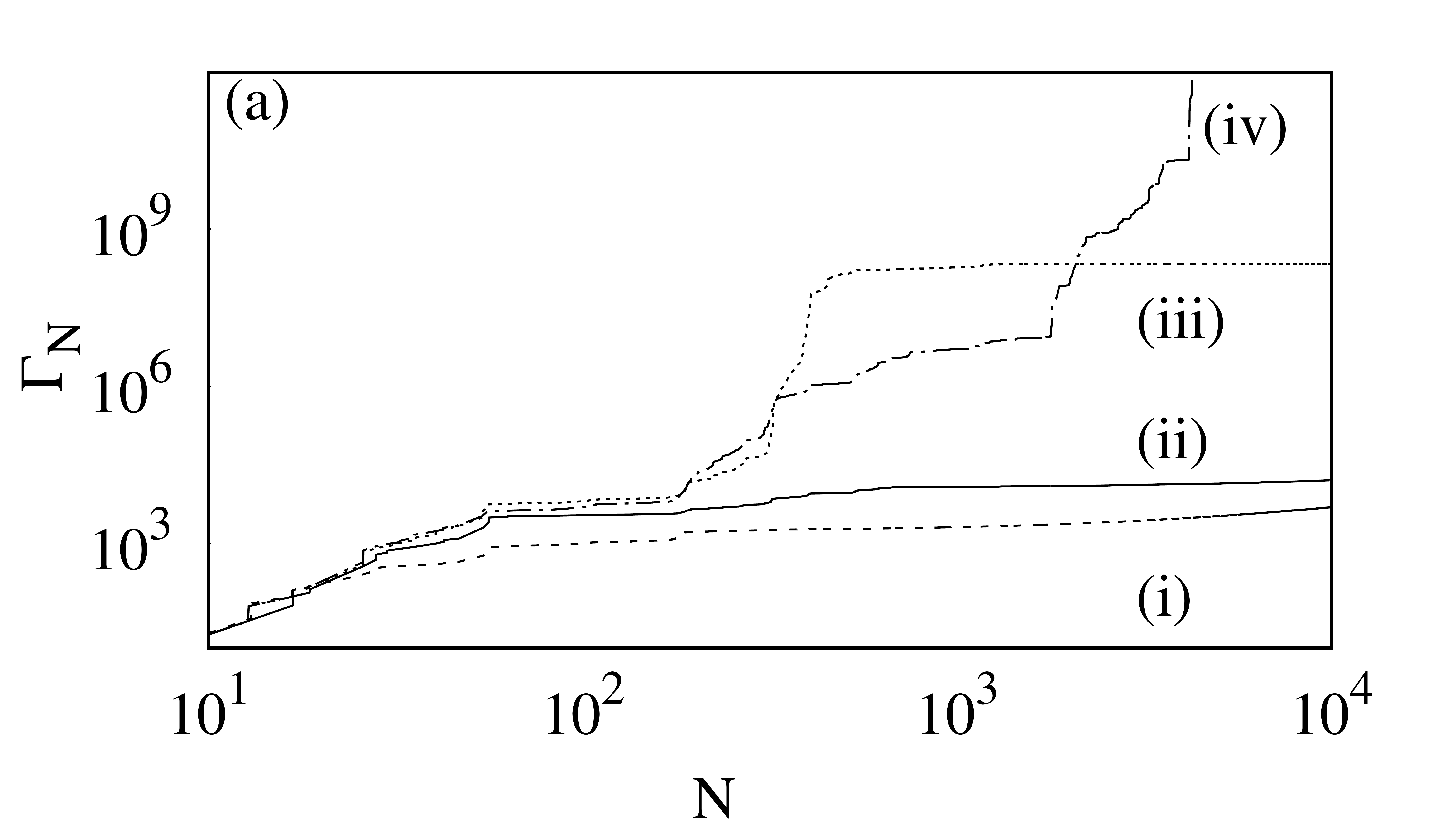} \\
	\includegraphics[width=0.9\linewidth]{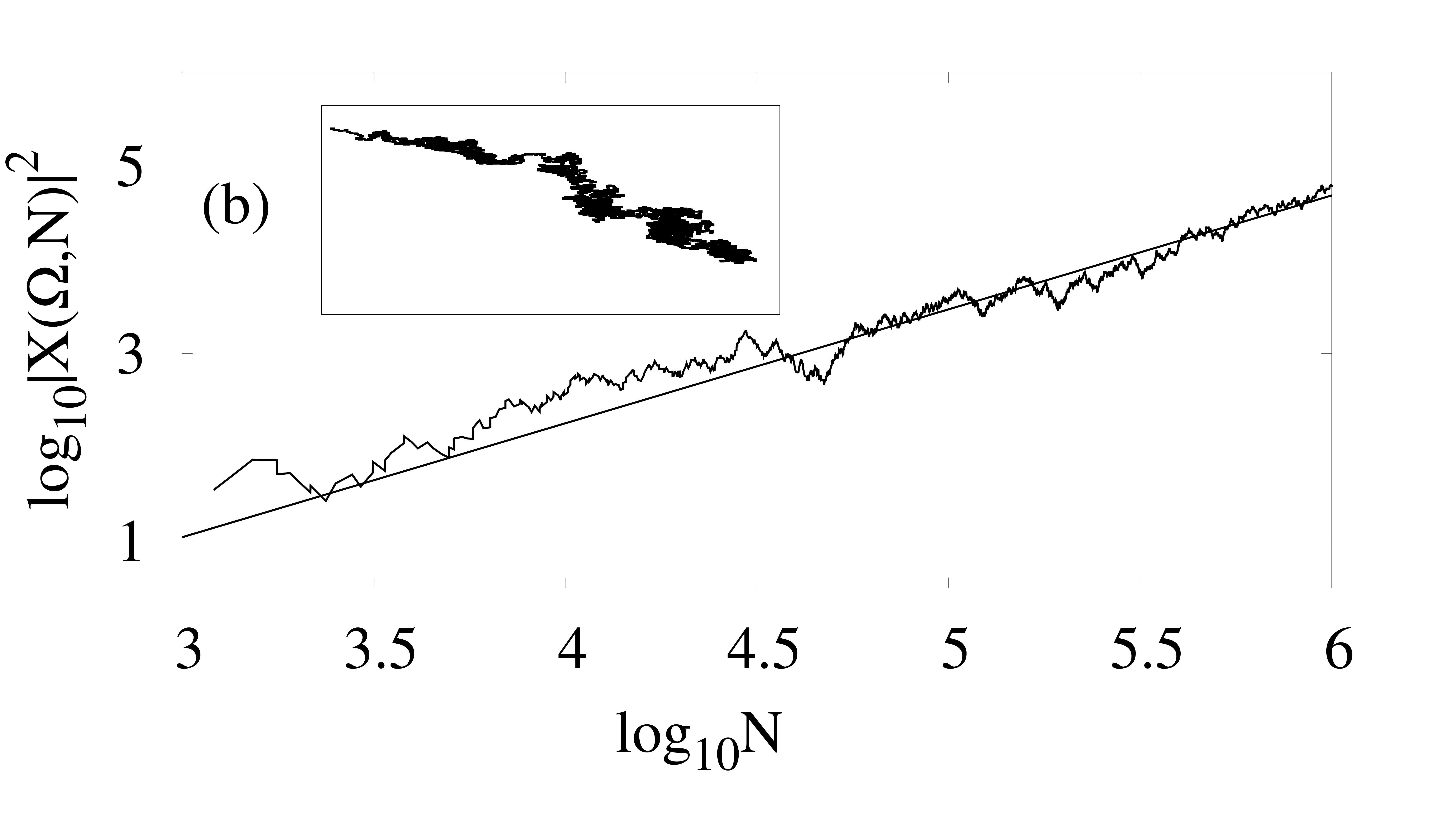}
	\caption{(a) Phase sensitivity exponent $\varGamma_{N}$ versus N showing (i) torus for A=0.3111, (ii) wrinkled torus for A=0.3112, (iii) SNA for A=0.311227 and (iv) chaos for A=0.31124.(b) Finite-time Fourier spectrum $|X(\Omega,N)|^{2}$ vs $N$ on logarithmic scale for SNA for A=0.311227 with the exponent $\beta=1.21$. The inset (b) shows fractal walk in the complex plane $(ReX,ImX)$ for the SNA attractor.}
	\label{fig3}
\end{figure}

Now we examine the dynamical transition to SNA by using specific quantities, namely Lyapunov exponents \cite{prasad2001strange}, phase sensitivity exponent and singular continuous spectrum \cite{ pikovsky1995singular, *pikovsky1995characterizing}. Fig.\ref{fig2} indicates the transition of the SNA into a chaotic attractor corresponding to a change in the largest Lyapunov exponent from negative to positive values at A=0.31124. To provide further evidence, we compute the phase sensitivity function $\varGamma_{N}$ which is bounded for torus region, grows with N as a power-law relation for SNA and  increases exponentially with N for chaotic oscillations, as shown in Fig.\ref{fig3}(a). Here, the phase sensitivity function is defined as $\varGamma_{N}=\underset{{x_{0},\phi_{0}}}{min} \bigg(\underset{0\leq n \leq N}{max}\bigg|\dfrac{dx_{n}}{d\phi}\bigg|\bigg)$. In addition, from the time-dependent Fourier transform $X(\Omega,N)=\sum\limits_{n=1}^{N}x_{n}e^{i2\pi n \Omega}$, for  $\Omega=\dfrac{\sqrt{5}-1}{2}$, the spectrum $|X(\Omega,N)|^{2} \sim N^{\beta}$ holds, and for SNA this scaling exponent takes the value $1<\beta<2.$ This behaviour is shown in Fig.\ref{fig3}(b) for SNA, where we observe a relatively robust power-law behaviour with $\beta=1.21$. It was also suggested that for SNAs, the spectral trajectory in the complex plane (Re X, Im X) should exhibit a fractal behaviour. This is indeed observed for the SNA attractor in this system as shown in the inset of Fig.\ref{fig3}(b).

\section{Effect of three-level square waves on the quasiperiodically driven Duffing oscillator}
Next, we analyse the response of the quasiperiodically driven nonlinear system \eqref{equ1} to deterministic logic input signal $I$, consisting of two square waves in the absence of noise. Specifically, for two logic inputs, we drive the system \eqref{equ1} with a low/moderate amplitude signal $I=I_{1}+I_{2}$ with two square waves of strengths $I_1$ and $I_2$ encoding two logic inputs. The inputs can be either 0 or 1, giving rise to four distinct logic input  sets $(I_1,I_2):(0,0),(0,1),(1,0)$ and $(1,1)$. For a logical  $'0'$, we set $I_{1}=I_{2}=-\delta$, whereas for a $'1'$, we  set $I_{1}=I_{2}=+\delta$, where $\delta$ represents a small/moderate  intensity input signal. Now the input sets (0,1) and (1,0) give rise to the same input signal $I$. As a result, the four distinct input combinations $(I_1,I_2)$ reduce to three distinct values of $I$, $-2\delta,~ 0,~ +2\delta$, corresponding to the logic inputs $(0,0),~(0,1)$ or $(1,0),~(1,1)$,  respectively. The output of the system is determined by the state $x(t)$ of system \eqref{equ1}; for example, if the output can be considered as logical $'1'$ if it is one particular state and logical $'0'$ if it is in a different state. Specifically, the output corresponding to this 2-input set $(I_1,I_2)$ for a system with the state values $x>0$ is taken to be $'1'$ and with $x<0$, it is taken to be $'0'$. So, when the system switches between these two states, the output toggles from logical $0$ to logical $1$ and vice-versa.

\begin{figure}
	\centering
	\includegraphics[width=0.9\linewidth]{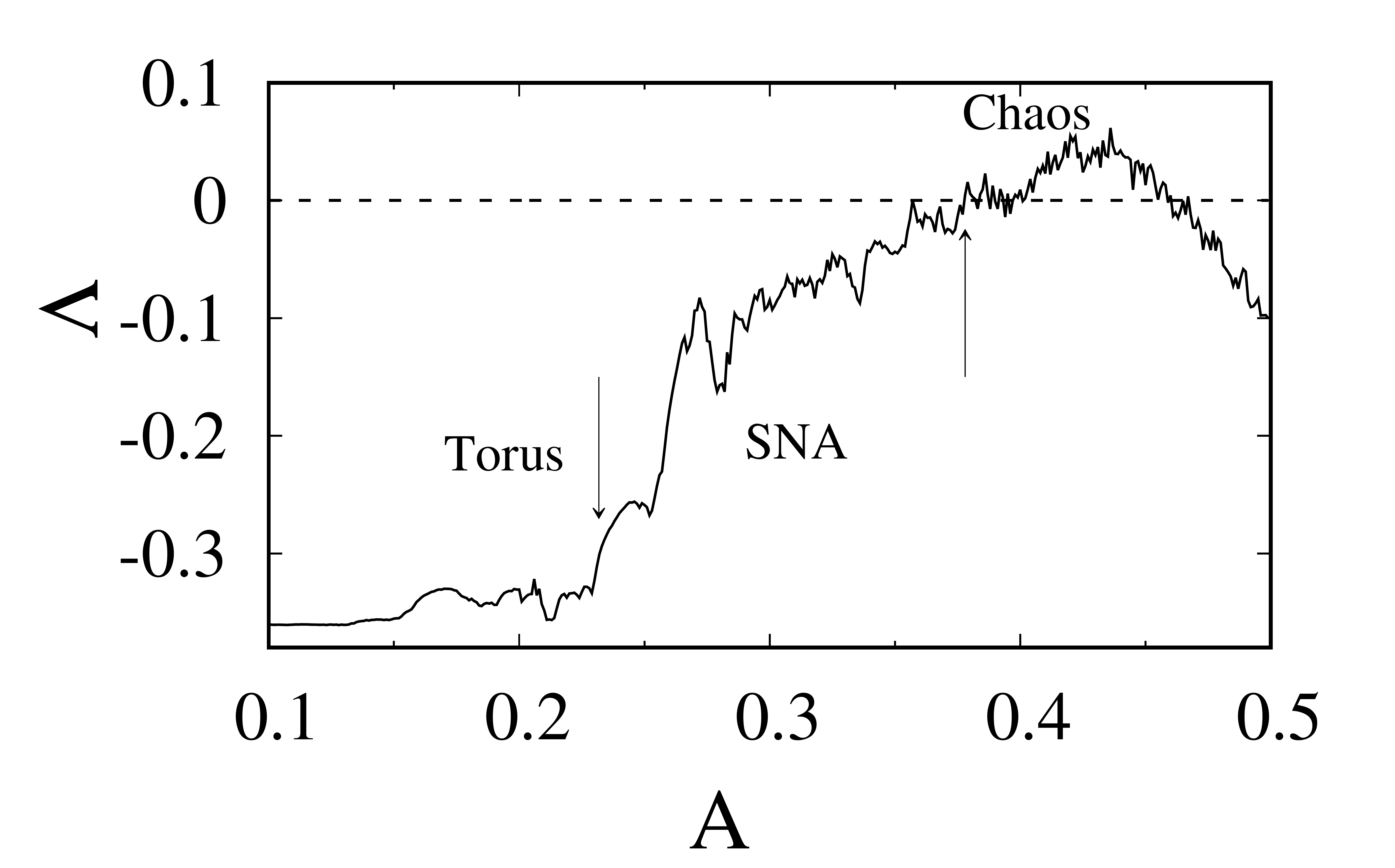}
	\caption{Maximal Lyapunov exponent  '$\Lambda$' versus the control parameter 'A': Solid curve corresponds to the system \eqref{equ1} after giving the inputs $I_{1}$, $I_{2}$ with bias $\varepsilon=0.05$, without noise (D=0.0).}
	\label{fig4}
\end{figure}
Here we will explicitly show that one indeed observes for a given set of inputs $(I_1,I_2)$ a logical output from the above quasiperiodically driven nonlinear system \eqref{equ1} in accordance with the truth table of the basic logic operations as given in Table \ref{Tab1}. 

\begin{table}
\begin{center}
\caption{Truth table of the basic logic operations} 
\vspace{0.2cm}
\begin{tabular} {|c| c| c| c| c|c|}
	\hline
	Input Set $I_{1}$, $I_{2}$ &OR & AND & NOR & NAND \\
	\hline
	(0,0)  & 0  & 0 &1&1 \\
	\hline
	(0,1)/(1,0)  & 1  & 0 &0&1  \\
	\hline
	(1,1)  & 1  & 1 &0&0  \\
	\hline

\end{tabular}
\label{Tab1}
\end{center}
\end{table}

\subsection*{A. Transition  to logical SNA}
\begin{figure}
\centering
\includegraphics[width=0.48\linewidth]{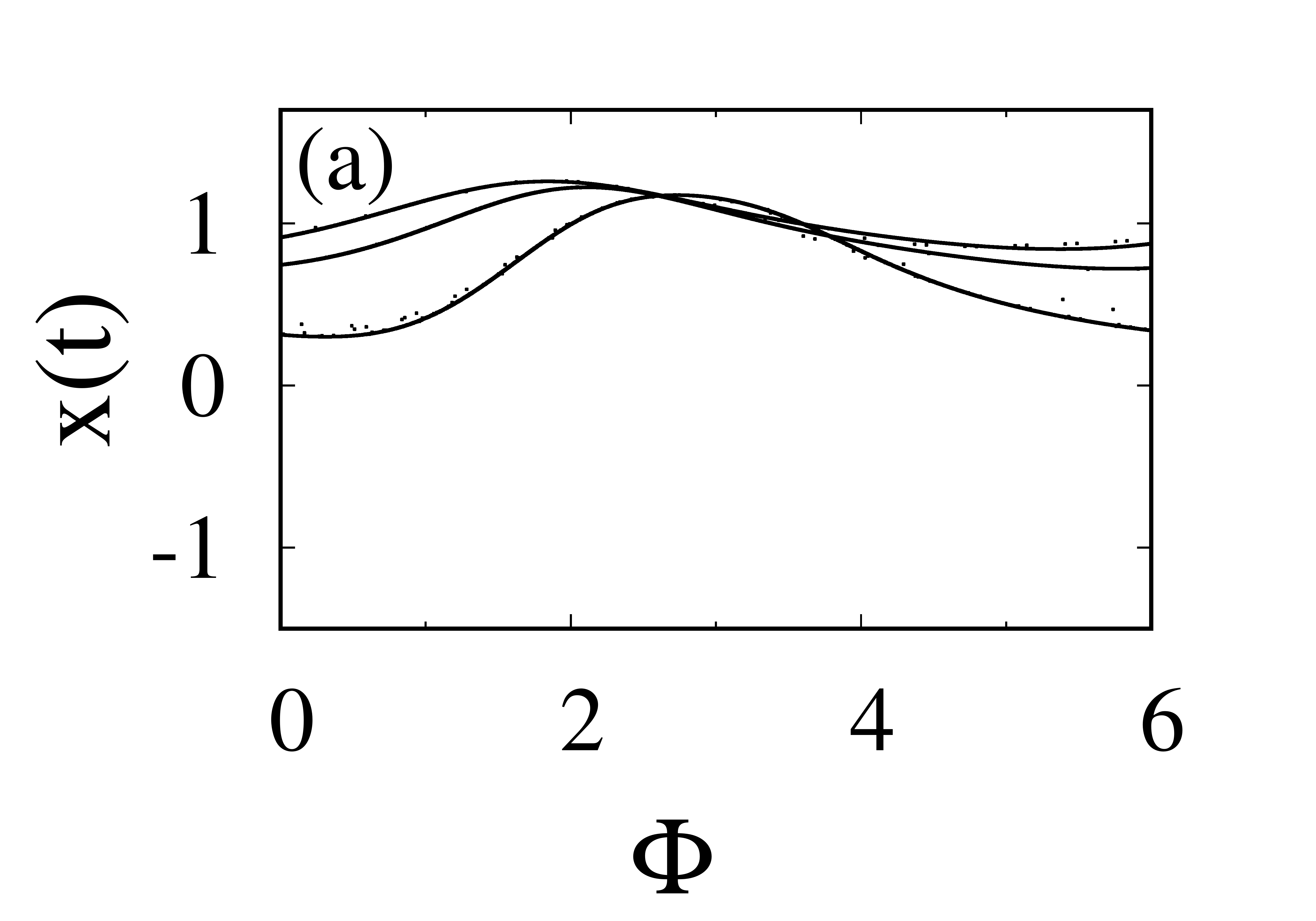}
\includegraphics[width=0.48\linewidth]{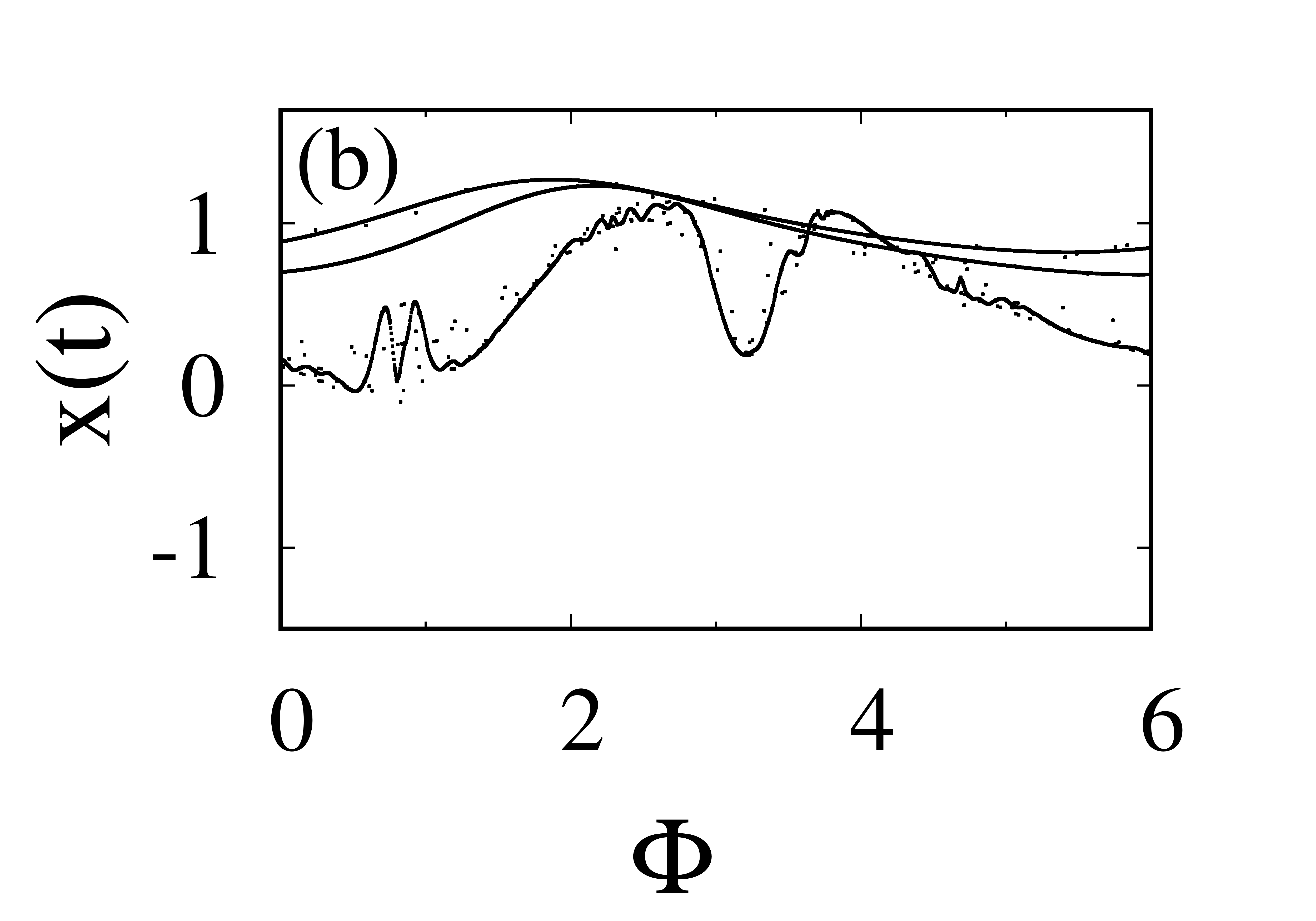}
\includegraphics[width=0.48\linewidth]{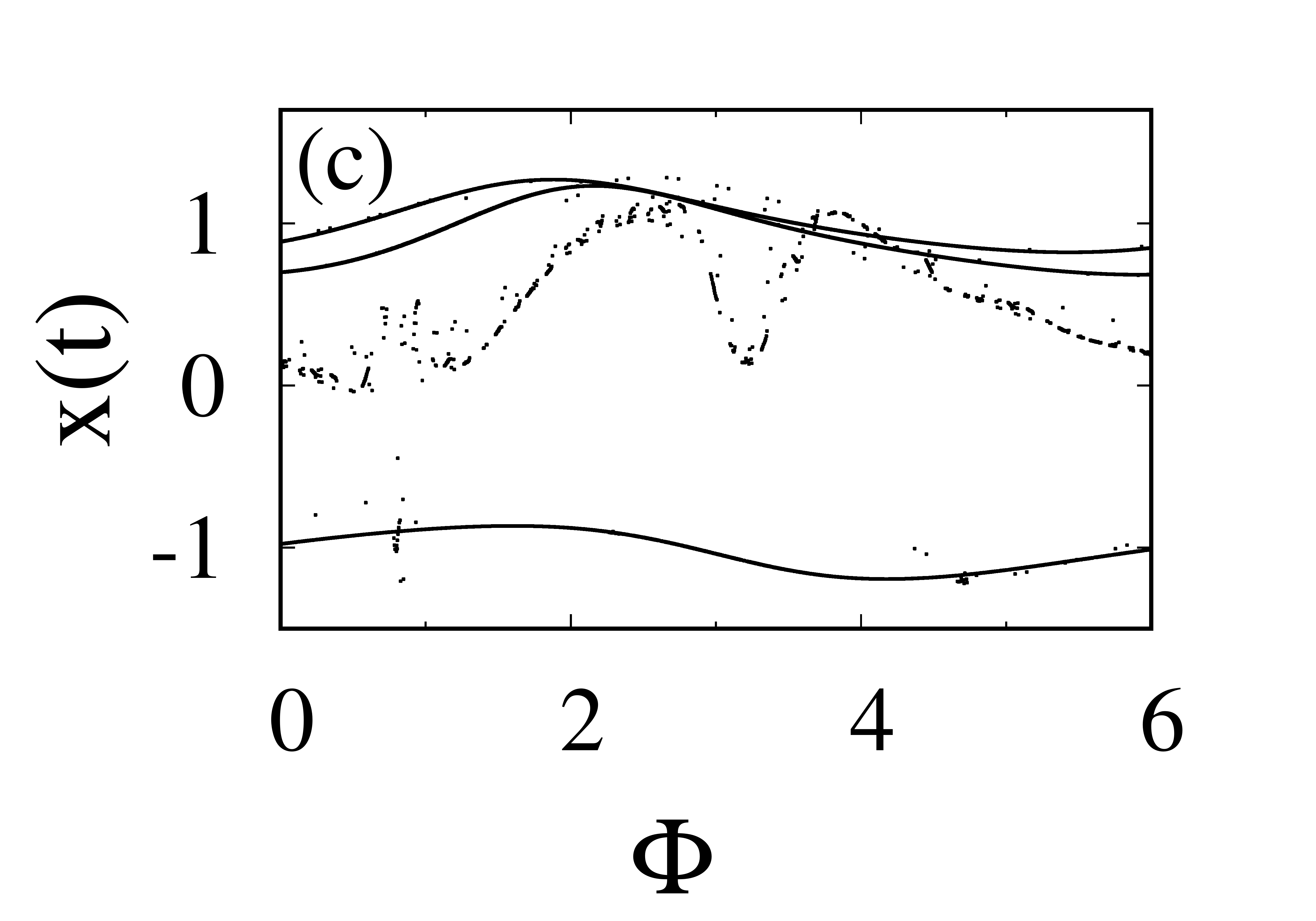} 
\includegraphics[width=0.48\linewidth]{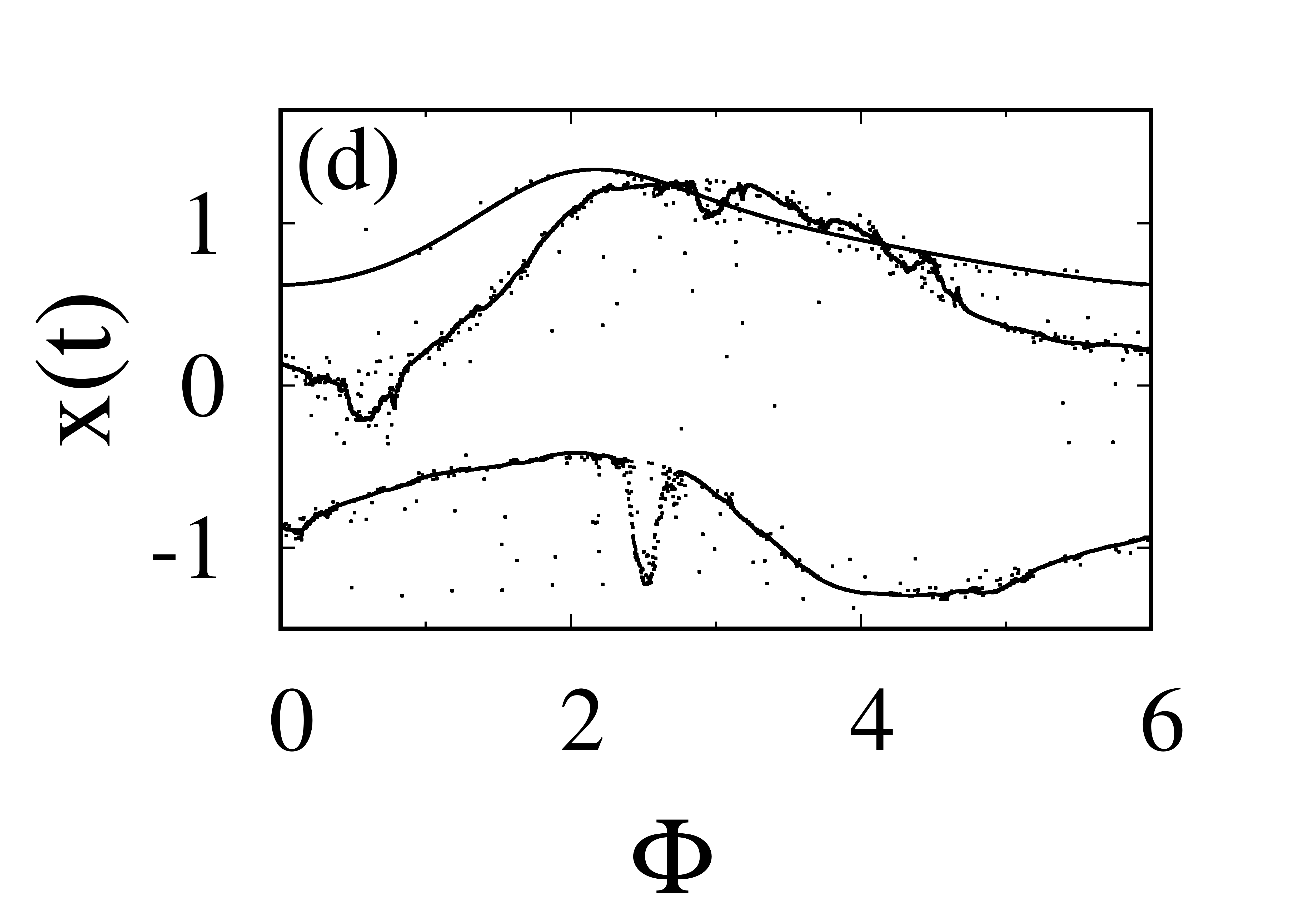}
\includegraphics[width=0.48\linewidth]{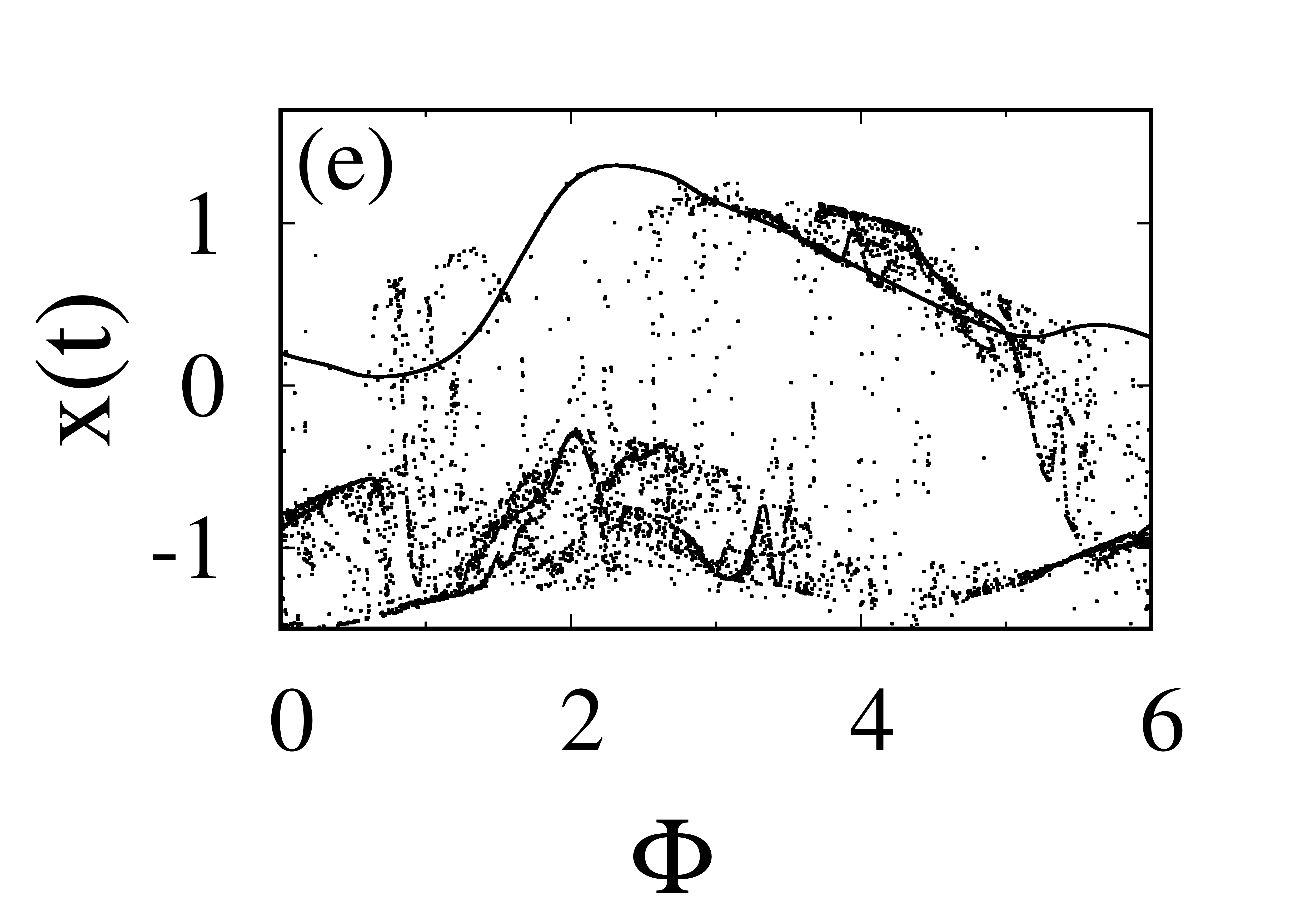}
\includegraphics[width=0.48\linewidth]{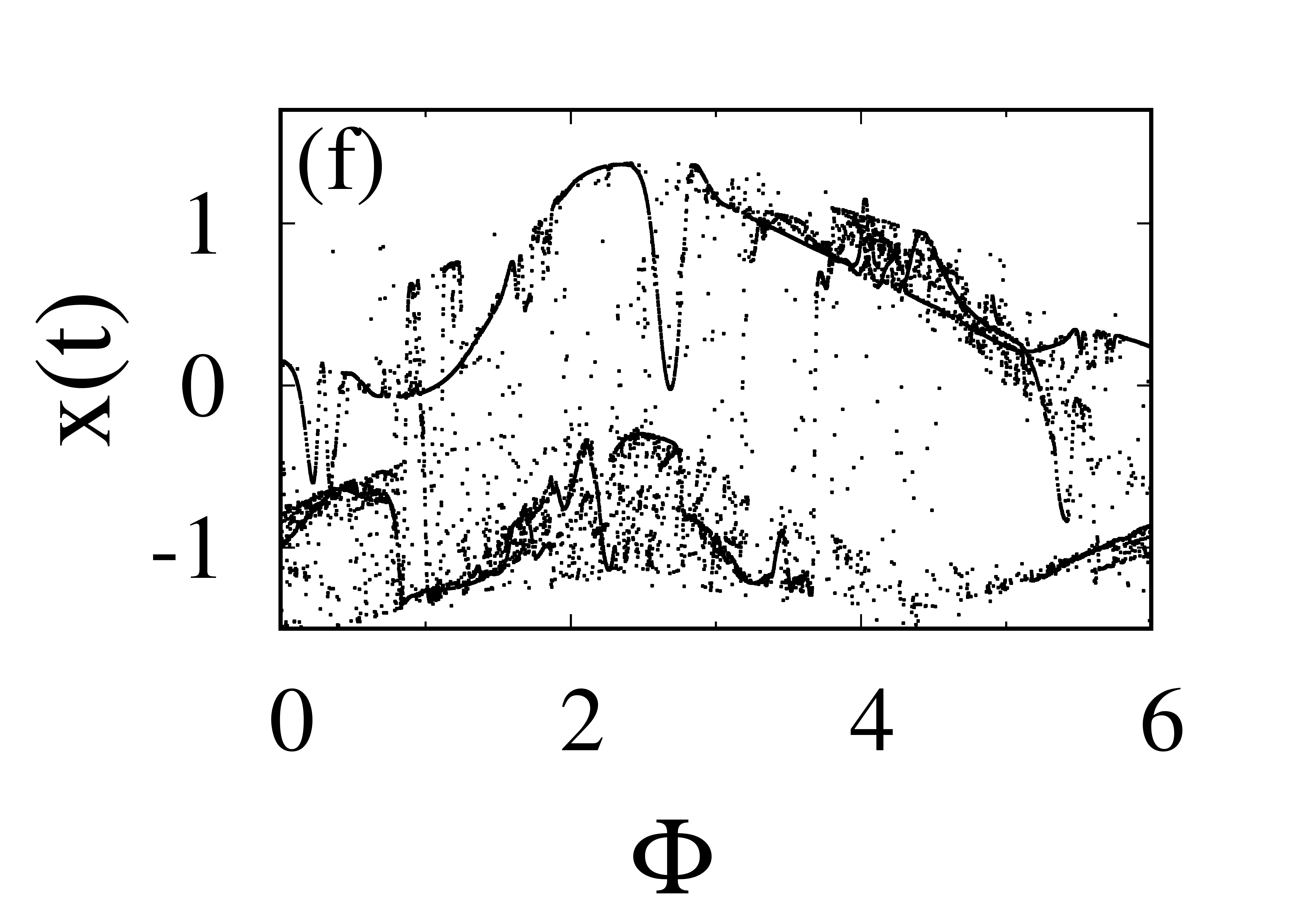}
\caption{Projection of the numerically simulated attractors of Eq.\eqref{equ1} in the ($\phi$ -x) plane for various values of 'A' (a) Period - 3 torus for A=0.22 (b) wrinkled torus for A= 0.23 (c) fractal torus for A= 0.2318 (d)  logical SNA for A=0.27 (e) standard SNA for A=0.32 (f) chaos for A = 0.392 and $\varepsilon$=0.05.}
\label{fig5}
\end{figure}

To be concrete, we consider the dynamics of (1) where both the inputs $I_1$ and $I_2$ take the values $-0.05$ when the logic input is $0$, and value $0.05$ when it is $'1'$.  The input signal $I=I_1+I_2$ is thus a three level square wave form: $-0.1$ corresponding to the input set $(0,0)$, $0$ corresponding to the input sets $(0,1)$ or $(1,0)$ and $0.1$ corresponding to the input set $(1,1)$. In our numerical experiments,  we choose $A$ as the bifurcation parameter and fix the other parameters at $\alpha=0.5, \beta=1.0,\omega_1=1.0,\omega_2 =\frac{1}{2} (\sqrt{5}-1)$, and $\varepsilon=0.05$. Fig.\ref{fig4} shows that with increasing forcing amplitude $A$ in \eqref{equ1}, the maximal Lyapunov exponent also grows and  that it changes sign (solid curve) at $A=0.383$. As the parameter $A$ is increased, we find that the transition from quasiperiodic orbit to chaos takes place in four distinct phases. In the first phase, a wrinkled torus becomes a  fractal torus. The second phase corresponds to a transition from the fractal torus to the logical SNA, where the basic logic operations become valid. The third phase is a transition from  the logical SNA to standard SNA, which does not exhibit logic operations. Finally a transition from SNA to a geometrically similar chaotic attractor occurs. These transitions are clearly seen in the numerical plots in the $(\phi-x)$ plane as shown in Fig \ref{fig5}.

\begin{figure}[!]
	\centering	
	\includegraphics[width=0.9\linewidth]{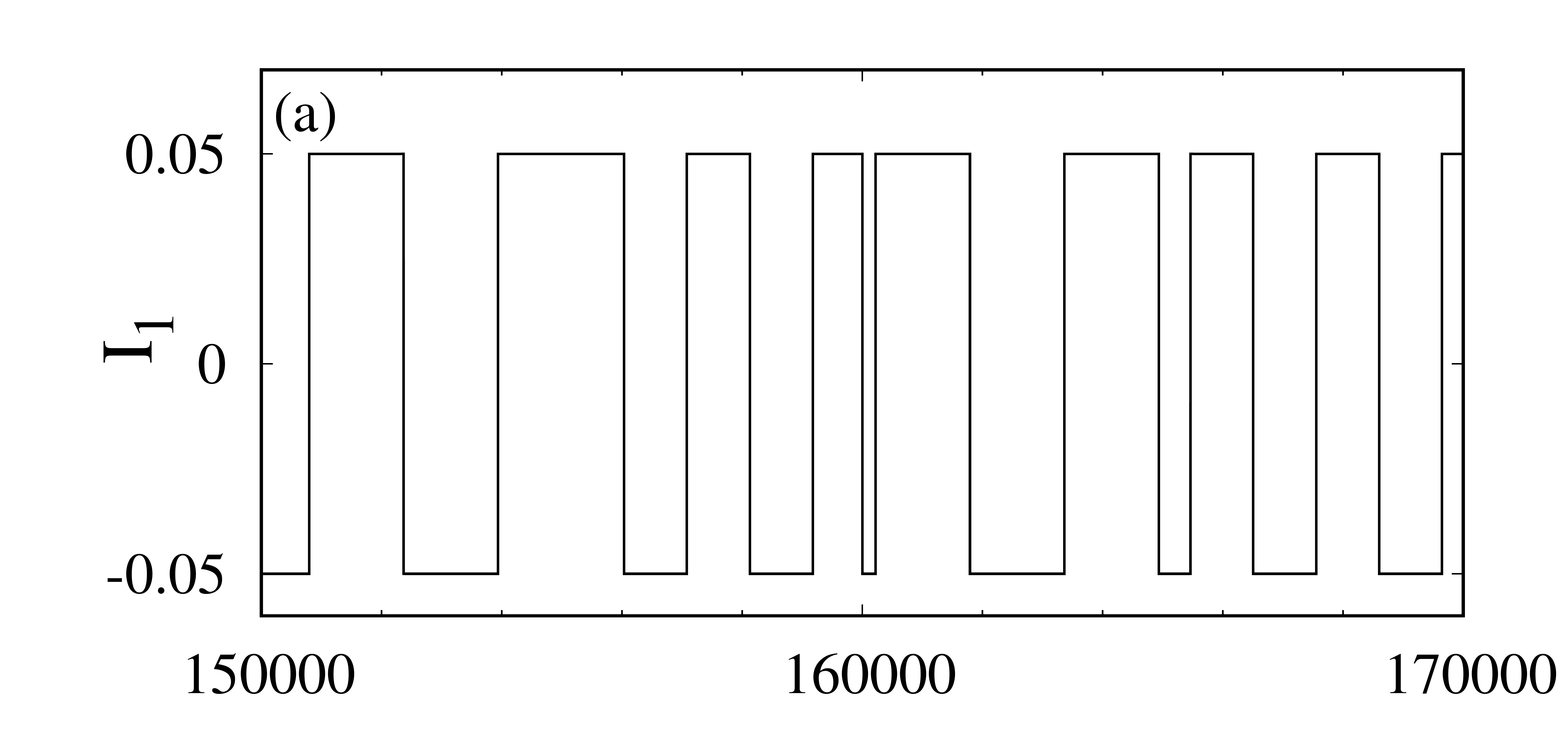} \\
	\includegraphics[width=0.9\linewidth]{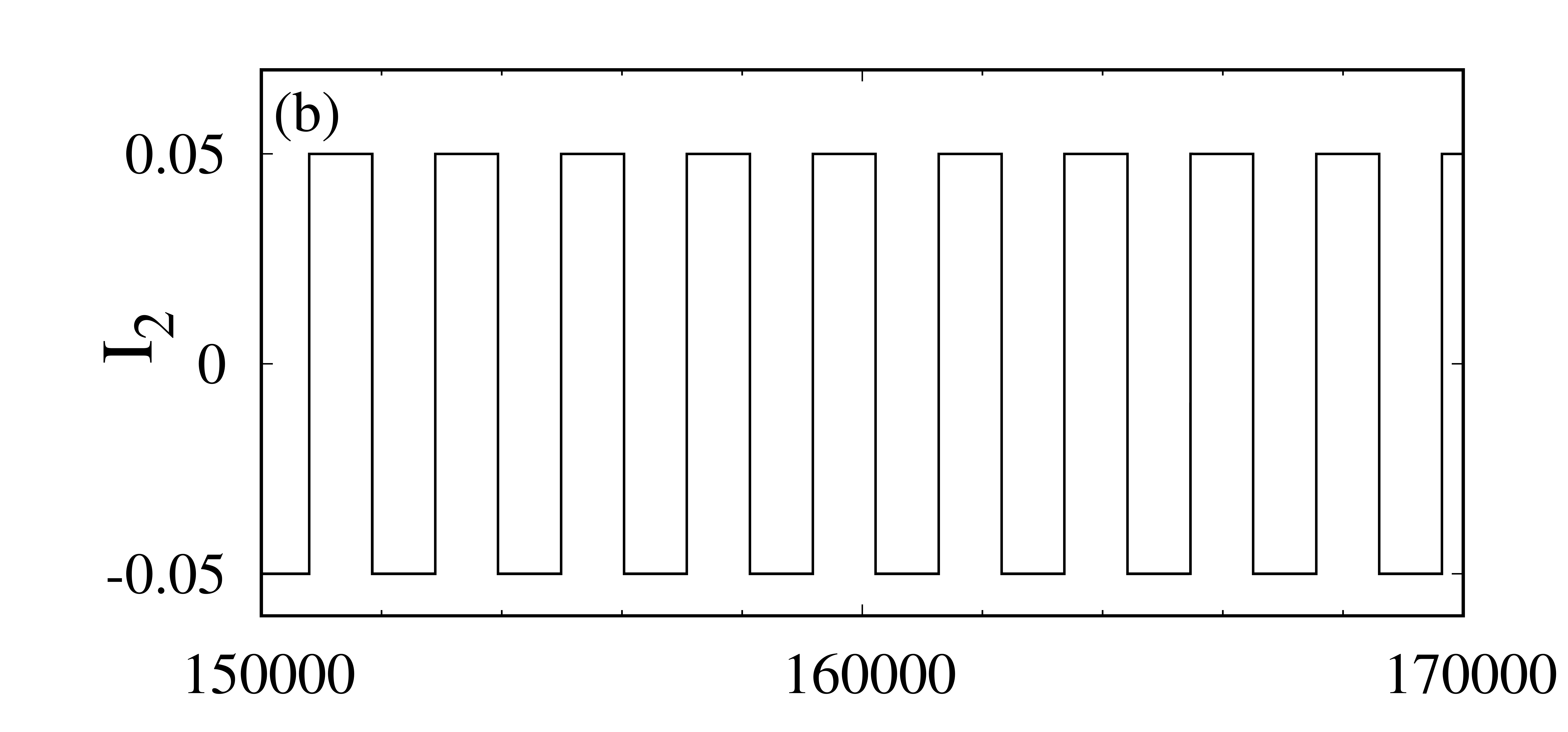} \\
	\includegraphics[width=0.9\linewidth]{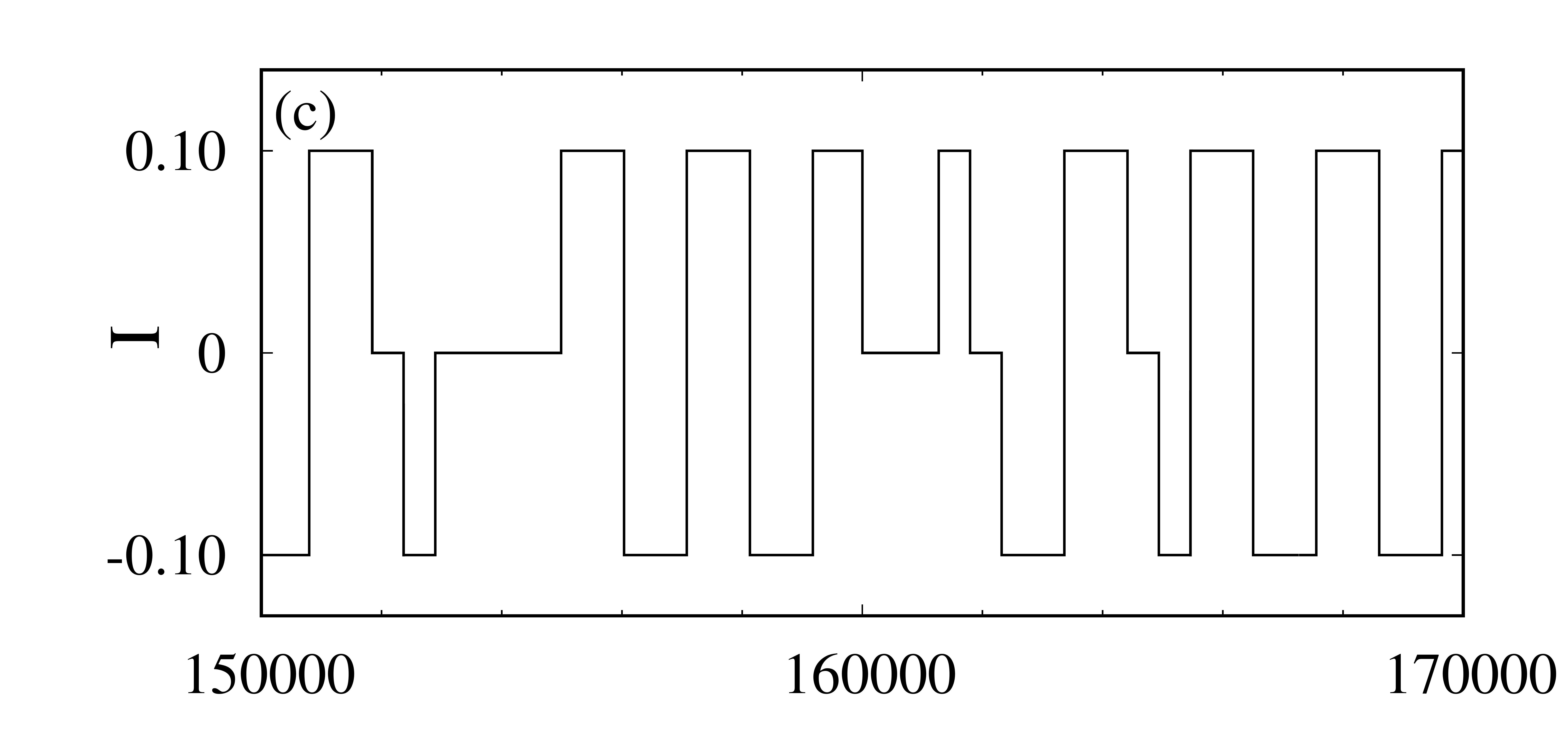} \\
	\includegraphics[width=0.9\linewidth]{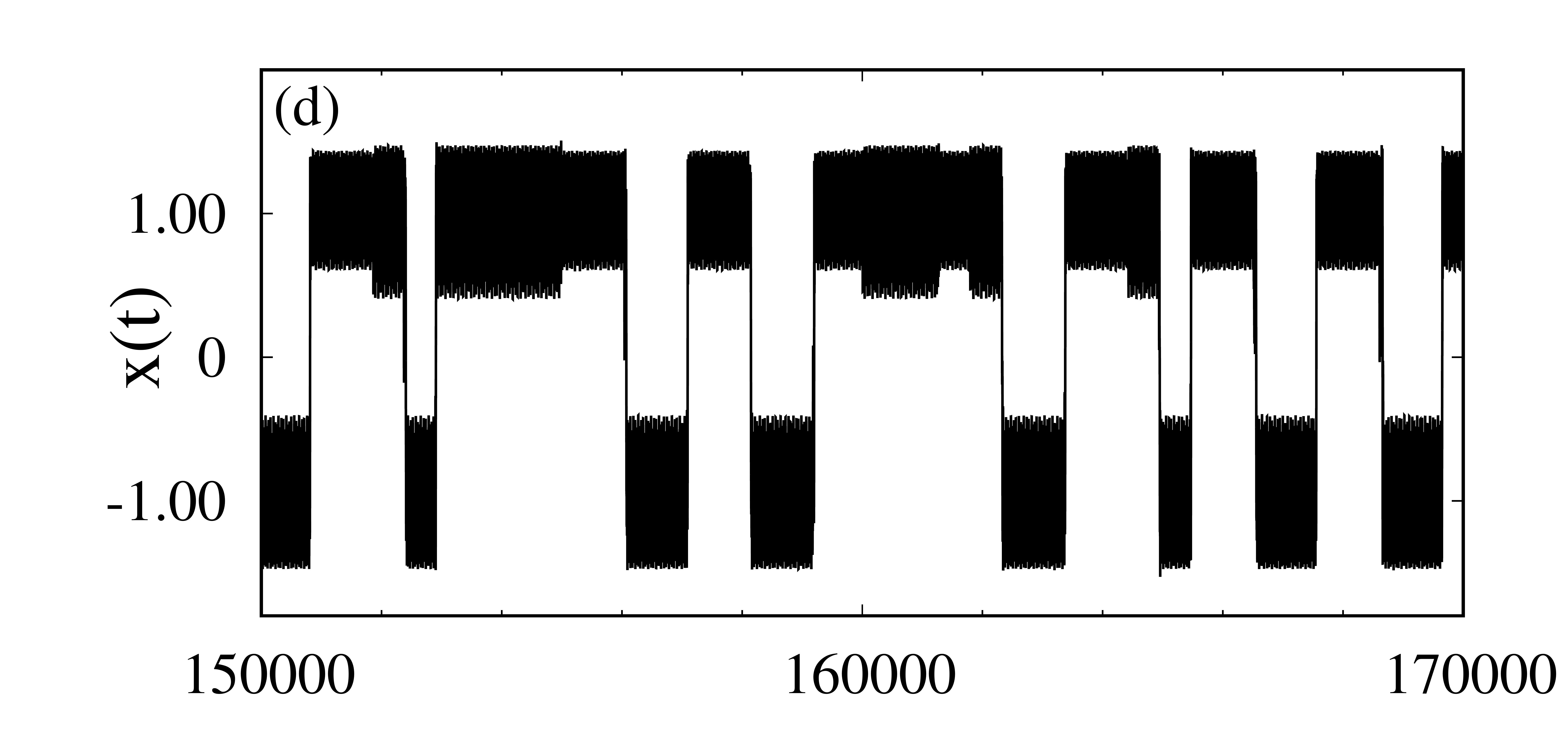} \\
	\includegraphics[width=0.9\linewidth]{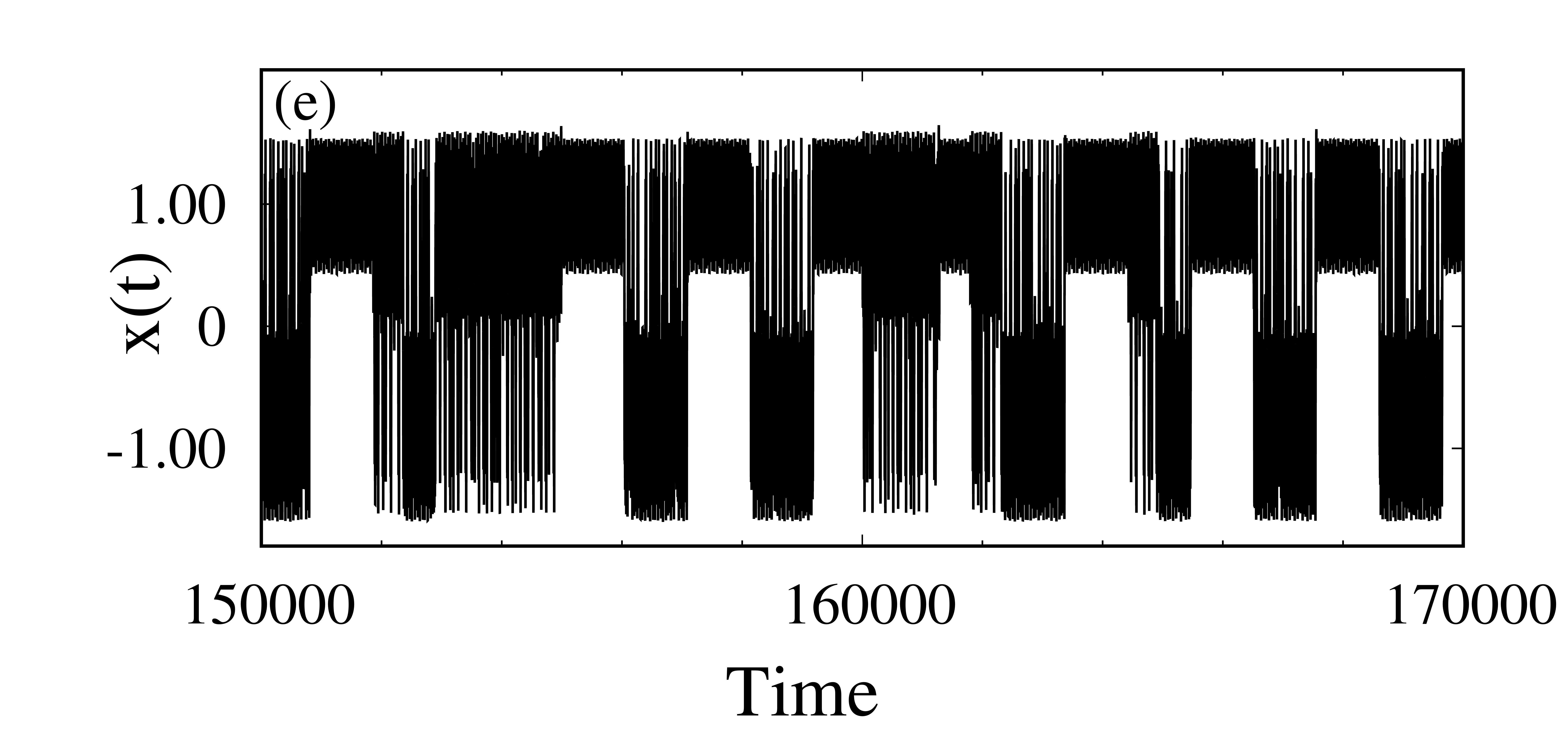}
	\caption{From top to bottom panels: (a)-(c) show a stream of input signals $I_{1}$, $I_{2}$ with $I_{1}=I_{2}=-0.05$ when the logic input is $'0'$ and $I_{1}=I_{2}=0.05$ when the logic input is $'1'$. The '3' level square waves with -0.1 corresponding to the input set (0,0), 0 for (0,1)/(1,0) set and 0.1 for (1,1) input set. Panels (d) and (e) represent the dynamical response of the system under quasiperiodic forcing for (i) A =0.27 and (ii) A = 0.32 respectively. Note that the quasiperiodic forcing is optimum when $A=0.27$  [panel(d)] where one obtains the desired OR logic outputs for $\varepsilon=0.05$ (see Table \ref{Tab1}).}
	\label{fig6}
\end{figure} 
\begin{figure}
	\includegraphics[width=0.48\linewidth]{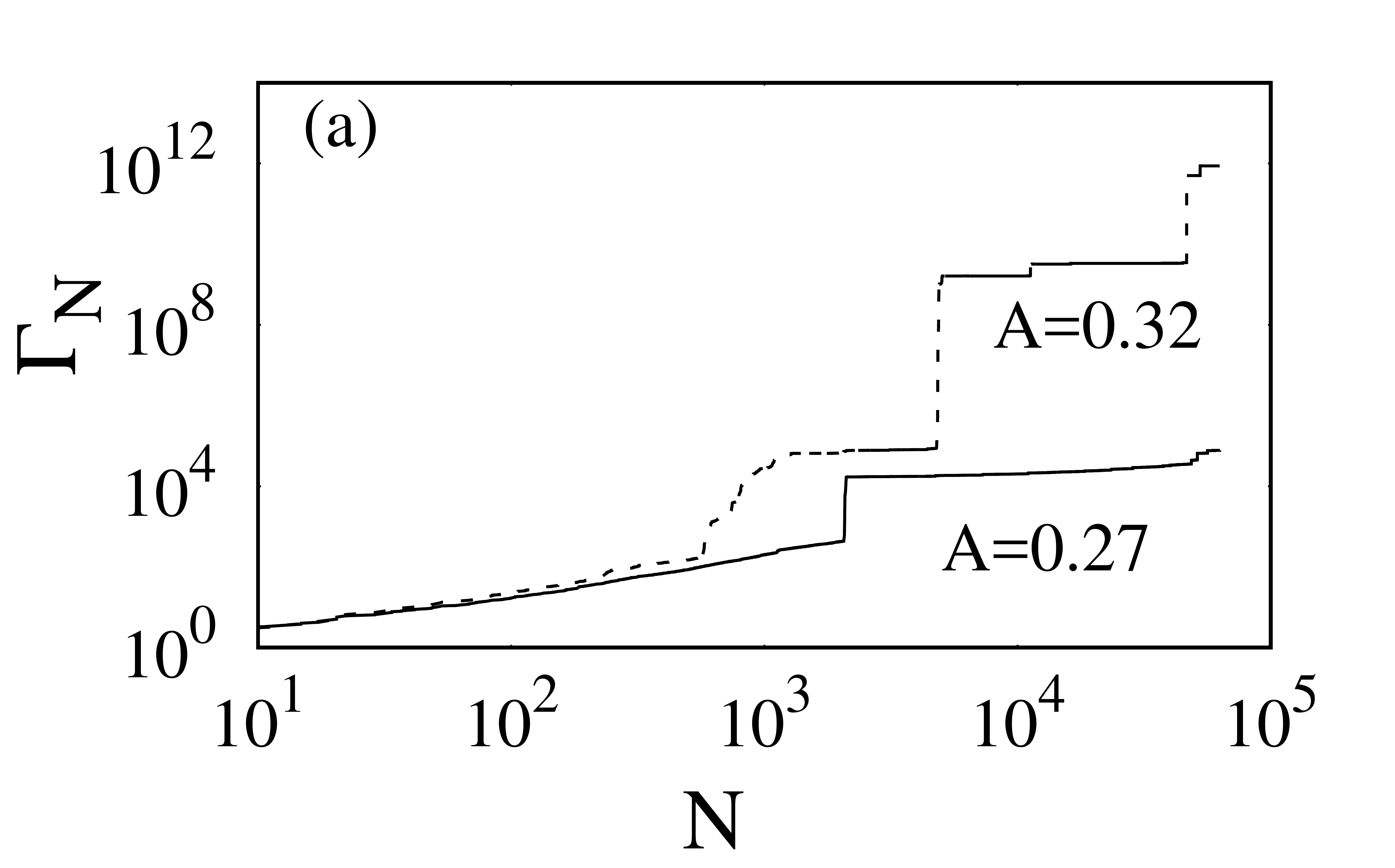}
	\includegraphics[width=0.48\linewidth]{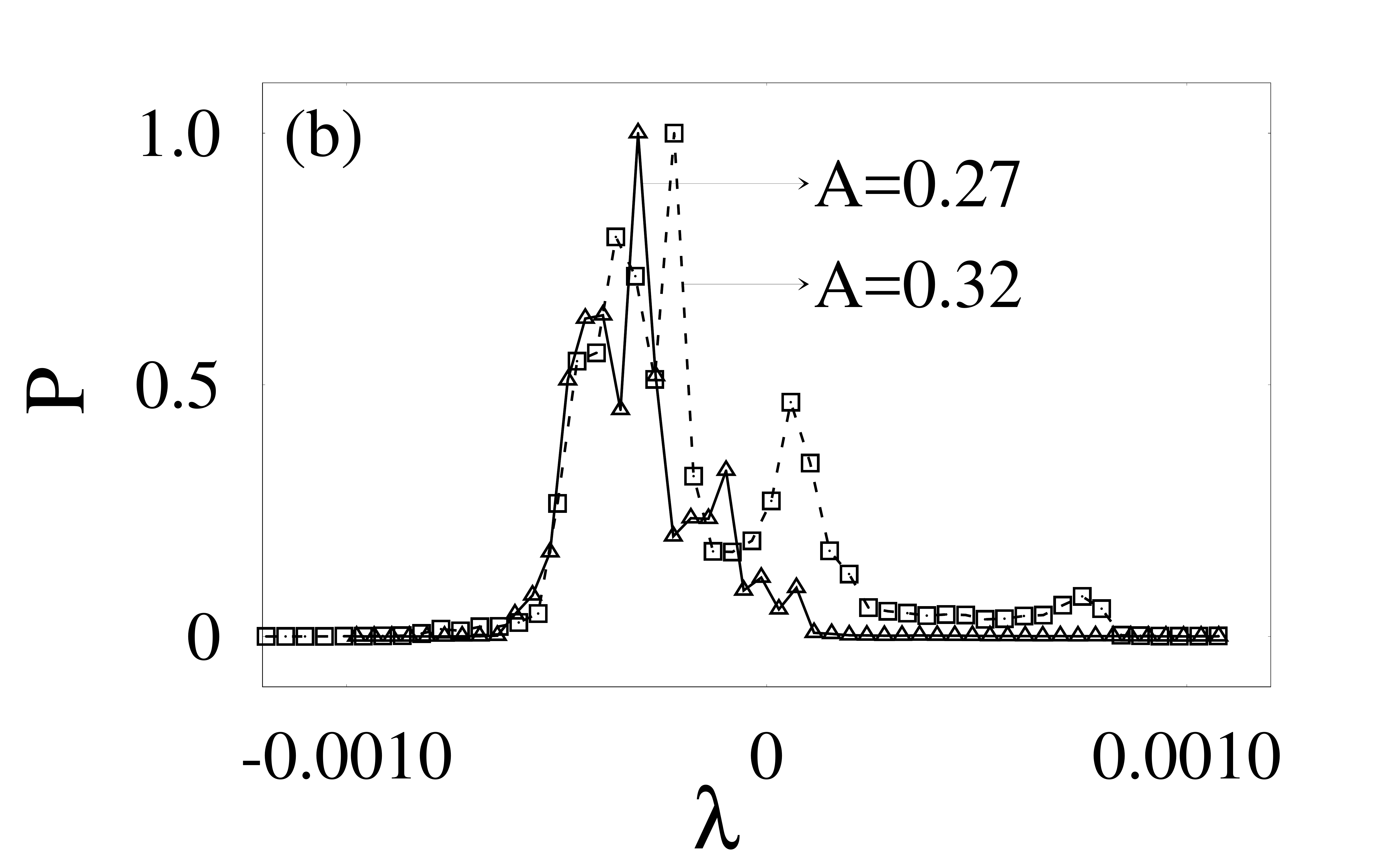} 
	\includegraphics[width=0.48\linewidth]{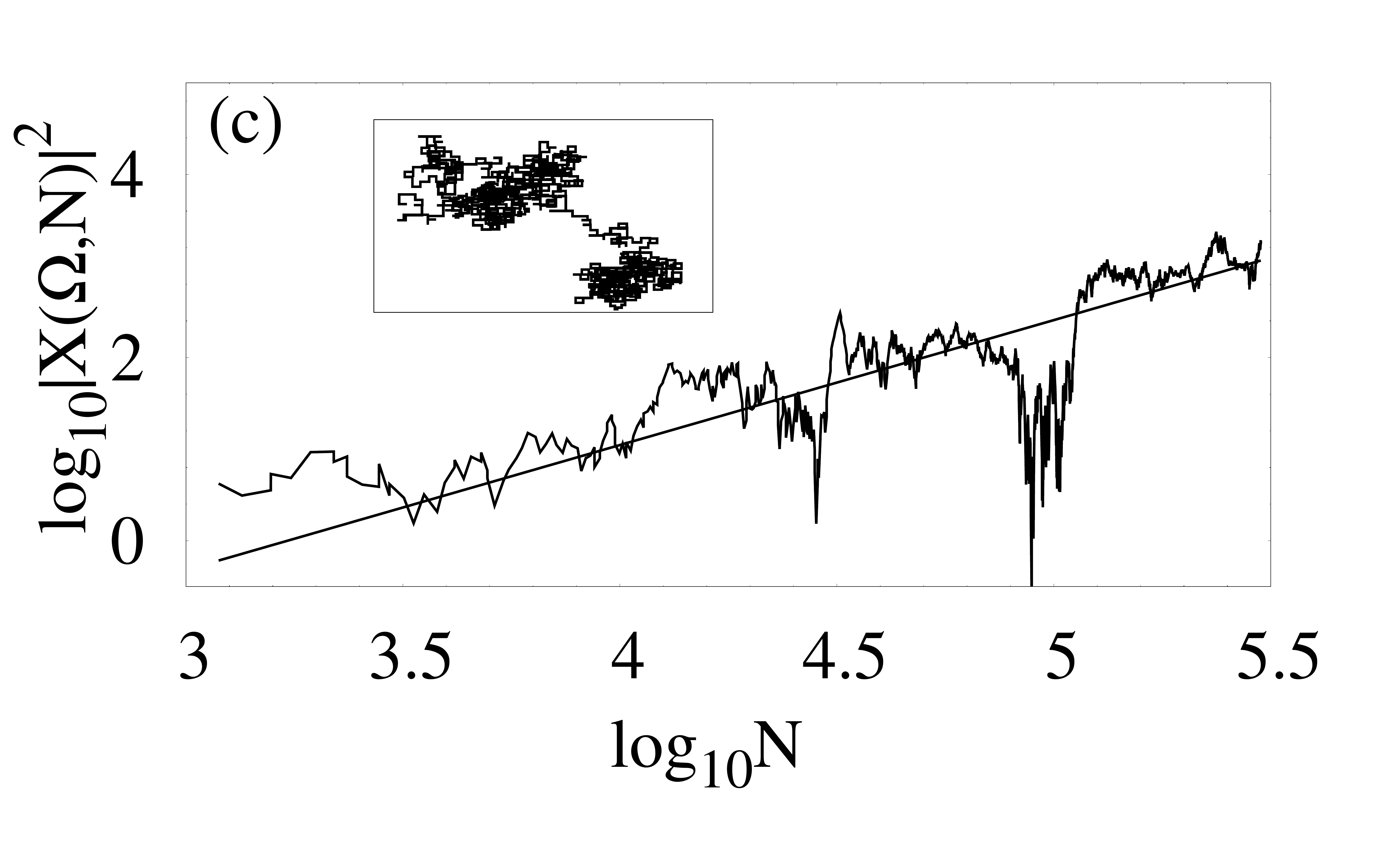}
	\includegraphics[width=0.48\linewidth]{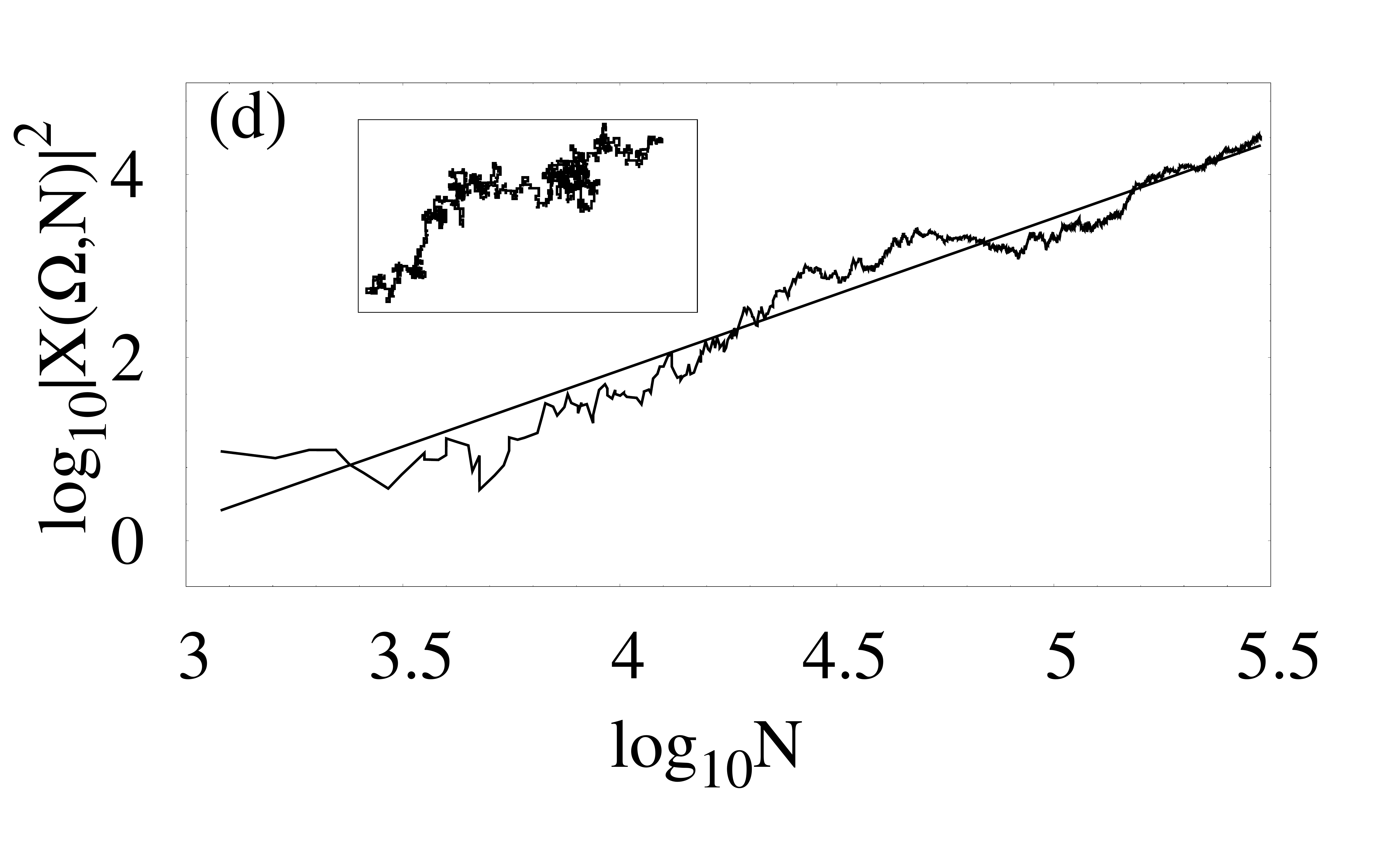}
	\caption{(a) Phase sensitivity exponent for $\varGamma_{N}$ versus N logical SNA for A=0.27 (continues line) and standard SNA for A=0.32 (dashed line), (b) Finite-time Lyapunov exponents for quasiperiodically driven Duffing oscillator including the effect of two aperiodic square waves: (i) logical SNA for A=0.27, (ii) standard SNA for A=0.32. Finite-time Fourier spectra $|X(\Omega,N)|^{2}$ vs $N^{\beta}$ on logarithmic scale for (c) logical SNA for A=0.27 with $\beta=1.3$, (d) standard SNA for A=0.32 with $\beta=1.6$. The insets in (c) \& (d) show a fractal walk in the complex plane $(ReX,ImX)$. .}
	\label{fig7}
\end{figure} 

 To confirm further, as the value of $A$ varies, a three-quasiperiodic torus is observed due to the effect of three level square waves and hence can be seen as three smooth branches in the Poincar\'e surface of section plot in the $(\phi-x)$ plane as shown in Fig.\ref{fig5}(a). This torus is either in $x>0$ well or $x<0$ well depending on the initial conditions we choose. In Fig.\ref{fig5}(a) we select the initial conditions such that the torus lies within the $x>0$ well. On increasing further the value of $A$ to $A=0.2232$ the torus begins to wrinkle. Fig.\ref{fig5}(b) reveals that among the three strands, one of the strands loses its smoothness and begins to wrinkle while other two stands are in the same well $(x>0)$.  These bends tend to become actual discontinuities at $A= 0.2312$.  At such values, the attractor loses its smoothness and becomes an SNA as the maximum Lyapunov exponent works out to be $\lambda=-0.032$. For such values of $A$, it is found that instead of the attractor bounded to a single well, it switches  between the two wells and can be seen in the Poincar\'e surface section as two strands in the $x>0$ well and one strand in the $x<0$ well as shown in Fig.\ref{fig5}(c). 

In particular, it is observed that the fractal torus involves a kind of sudden widening of the attractor similar to the crisis phenomenon that occurs in chaotic systems. It is seen in the $(\phi-x)$ plane as shown in Fig.\ref{fig5}(d) for A=0.27, the orbit in the attractor spends long stretches of time in the region at which the attractor is confined to a particular well ($x>0$). At the end of these long stretches, the orbit switches out of the well and spends around the other well ($x<0$) due to crisis. It then returns to the old region for another stretch of time, followed by a burst and so on. This kind of widening of the attractor usually occurs in chaotic system at a crisis.

However, in the present case, we have shown the existence of such a possibility in a quasiperiodically forced system by creating SNAs. It is very clear from these transitions that the SNA created via fractalization becomes a logical SNA through widening crisis. This kind of SNA is due to the aperiodic input signals. That is, in an optimal range of $A$, $0.23<A<0.31$, the output of the system synchronizes with the aperiodic input signal. If the aperiodic input signal follows any kind of logic behaviour, the response of the system also follows the same. Therefore this kind of attractor can be called logical SNA. On further increase in the value of $A$, $0.311<A<0.3823$, the logical SNA loses its synchrony with the input signal and becomes the standard SNA. This can be seen in the $(\phi-x)$ plane when all the strands lose their continuity as shown in Fig.\ref{fig5}(e). Increasing the value further, $A>0.3824$, the attractor becomes chaotic with a geometry similar to SNA [see Fig.\ref{fig5}(f)].

Specifically, we note that when the input signals $(I_1,I_2)$ are in the $(1,1)$ or $(0,1)/(1,0)$ states the attractor resides  in the $x>0$ well and when the input signal is in the $(0,0)$ state, the attractor  is in the $x<0$ well. We observe that under optimal quasiperiodic forcing strength $0.23 < A< 0.31$, the state $x<0$ as logic output $0$ and the state $x>0$ as logic output $1$ yield a clean stable logical OR gate SNA with $\varepsilon=0.05$ [see Fig.\ref{fig6} for full details]. In a completely analogous way if we interpret the state $x<0$ as logic output $1$ and the state $x>0$ as logic output $0$, one realizes a stable logic NOR gate. Similarly, AND and NAND gates can be identified for a different value of bias $\varepsilon$ as shown in the next section.

On further increase in the value of $A>0.311$ the orbits due to the $(0,0)$ state and due to the $(1,0)/(0,1)$ state cause the attractor to wrinkle as seen in Fig.\ref{fig5}(e) and become standard SNA, where the logic no longer works. Here both the logical SNA and standard SNA are confirmed by computation of the largest Lyapunov exponents, which show that these attractors [Figs.\ref{fig5}(d), \ref{fig5}(e) \& \ref{fig6}(d), \ref{fig6}(e) ] are strange and nonchaotic.

\subsection*{B. Mechanism for logical strange nonchaotic attractor}
Let us now point out the mechanism of logical strange nonchaotic attractors. When a weak/moderate aperiodic three level square wave signal is applied to a bistable system, it serves to aperiodically modulate the potential by raising and lowering the wells. Essentially, the additive forcing changes erratically the relative depth of potential wells, thereby increasing the probability of jumps between wells. At a critical value of the quasiperiodic forcing, the particle in a well arrives at the neighboring of the barrier so that the quasiperiodic forcing is able to push it to the other well. At this junction, the system output $x(t)$  [see Fig.\ref{fig6}(d)] attains the same behaviour as the three level square waves [see Fig.\ref{fig6}(c)]. 
The essential ingredients for this behaviour consists of a nonlinear system, a three-level square wave and a source of quasiperiodic forcing. Further increase of quasiperiodic forcing produces a loss of coherence between $x(t)$ and $I$ [see Fig.\ref{fig6}(e)]. For sufficiently large value of 'A', the motion is strongly dominated by the quasiperiodic forcing. In this the intermittent dynamics disappears, and the trajectory jumps erratically between the wells,  [see Fig.\ref{fig5}(e)] and the dynamics still persists as strange nonchaotic. 

\subsection*{C. Characterization of logical SNA}
The phase sensitivity exponent $\varGamma_{N}$ for the quasiperiodic force for A=0.27 and A=0.32 are obtained for logical and standard SNAs as shown in Fig.\ref{fig7}(a). It  grows with N with a kind of power-law relation for the SNA. The distributions of finite time Lyapunov exponents for logical SNA and standard SNA are shown in Fig.\ref{fig7}(b). Both cases exhibit stretched exponential tails. From the spectral properties it is evident that for both logical SNA and standard SNA, the power spectrum varies as $|X(\Omega,N)|^{2} \sim N^{\beta}$,  where $\beta=1.3$ for logical SNA and $\beta=1.6$ for standard SNA, Figs.\ref{fig7}(c) and Figs.\ref{fig7}(d) respectively. The insets in these figures also demonstrate the fractal walk of the trajectories in the complex  (ReX, Imx) plane, as required for SNAs.

\subsection*{D. Transition to logical SNA : Nature of physical relationship between input and output responses}

\begin{figure}
	\centering
	\includegraphics[width=0.9\linewidth]{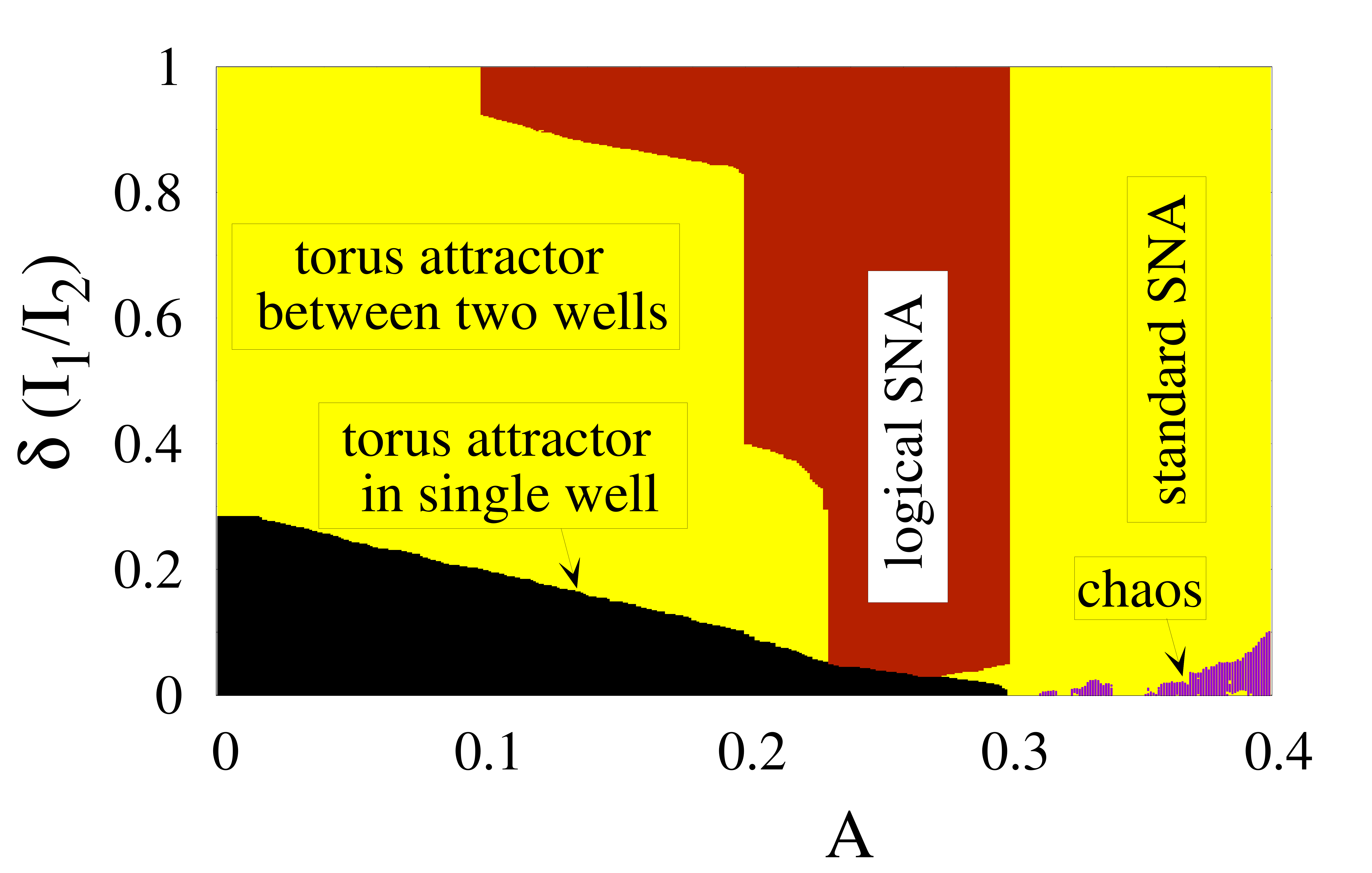}
	\caption{Two parameter phase diagram for strength A versus amplitude $\delta$: torus attractor in single well or double well, logical SNA, standard SNA and chaos as indicated in the figure.}
	\label{fig8}
\end{figure} 

\begin{figure}
	\centering	
	\includegraphics[width=0.9\linewidth]{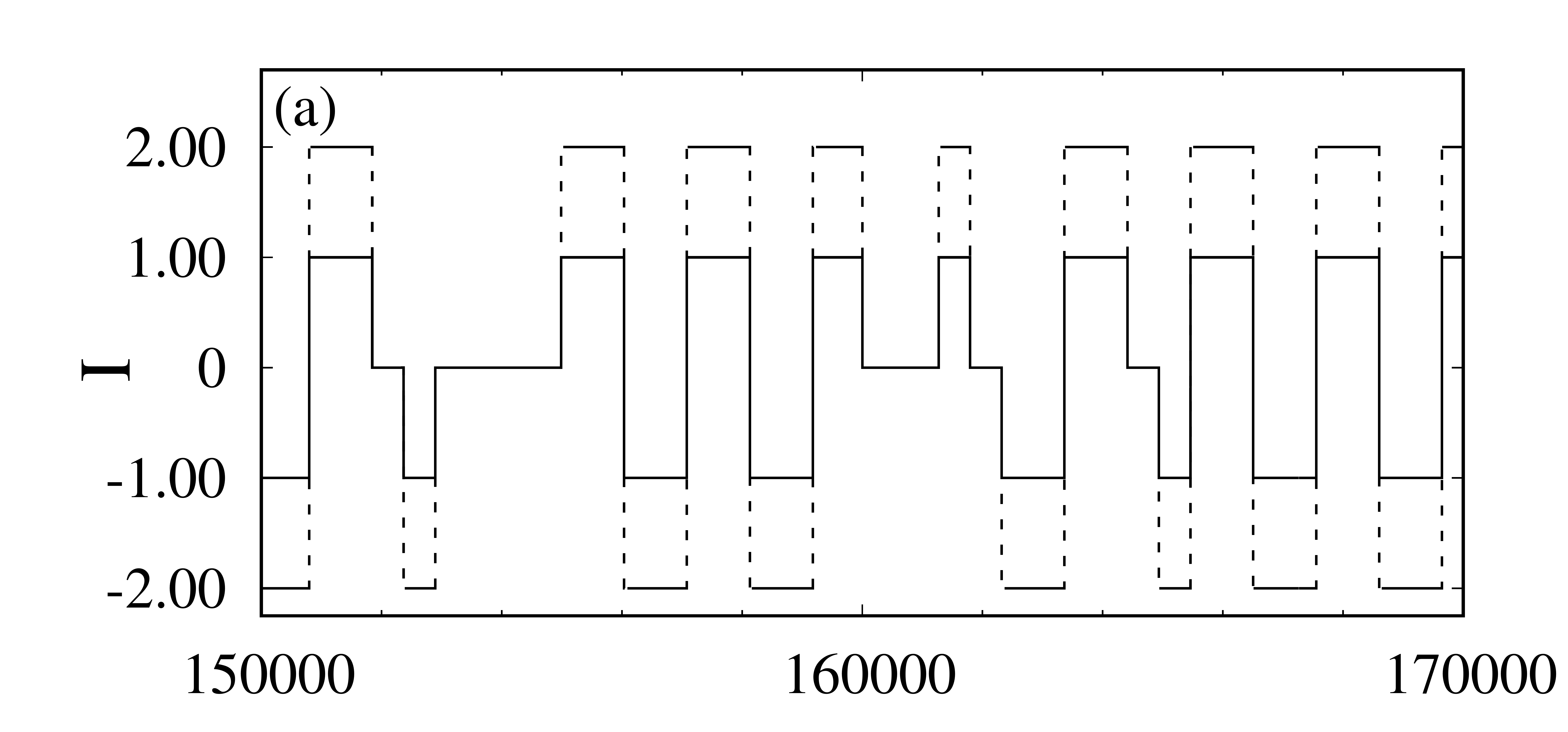} 
	\includegraphics[width=0.9\linewidth]{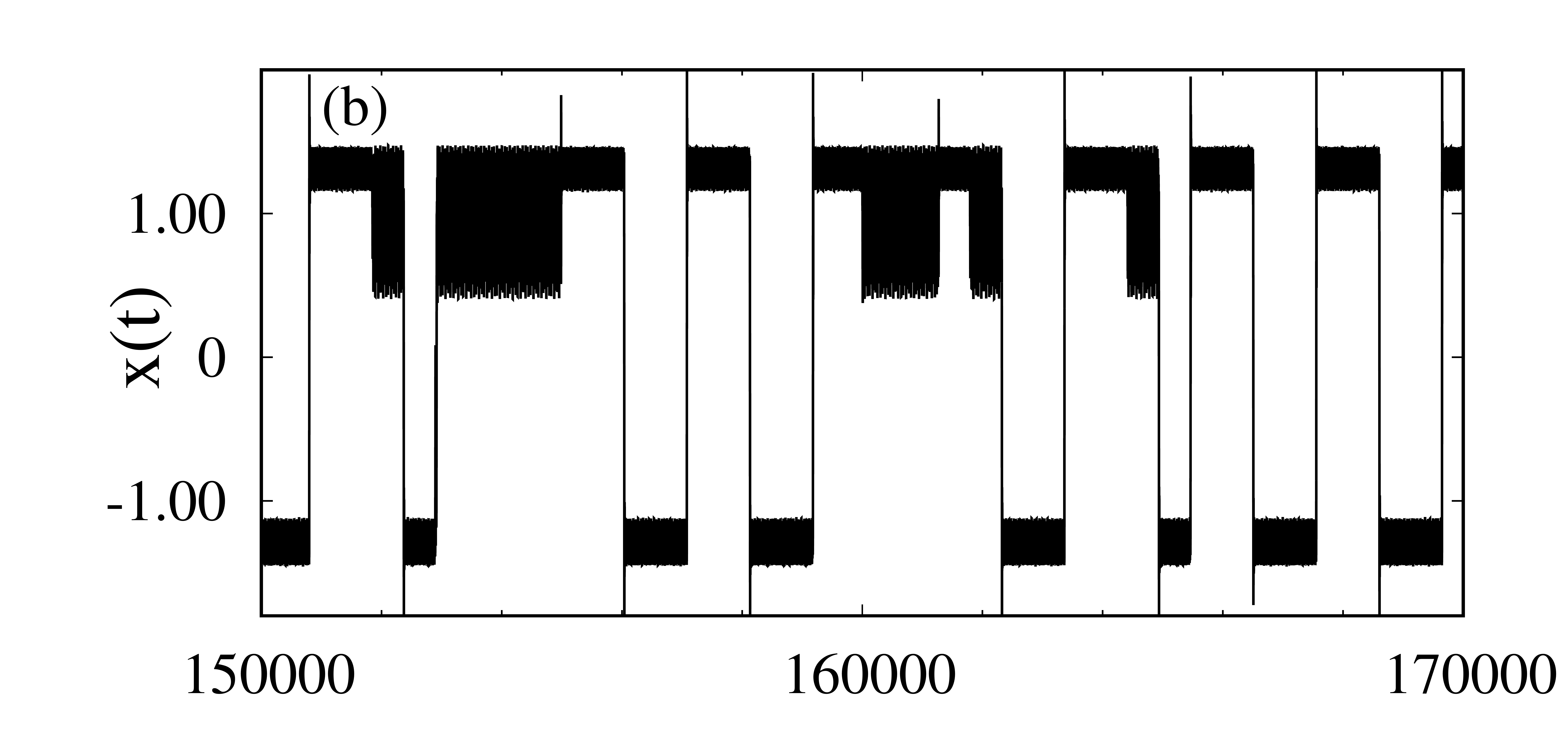} \\
	\includegraphics[width=0.9\linewidth]{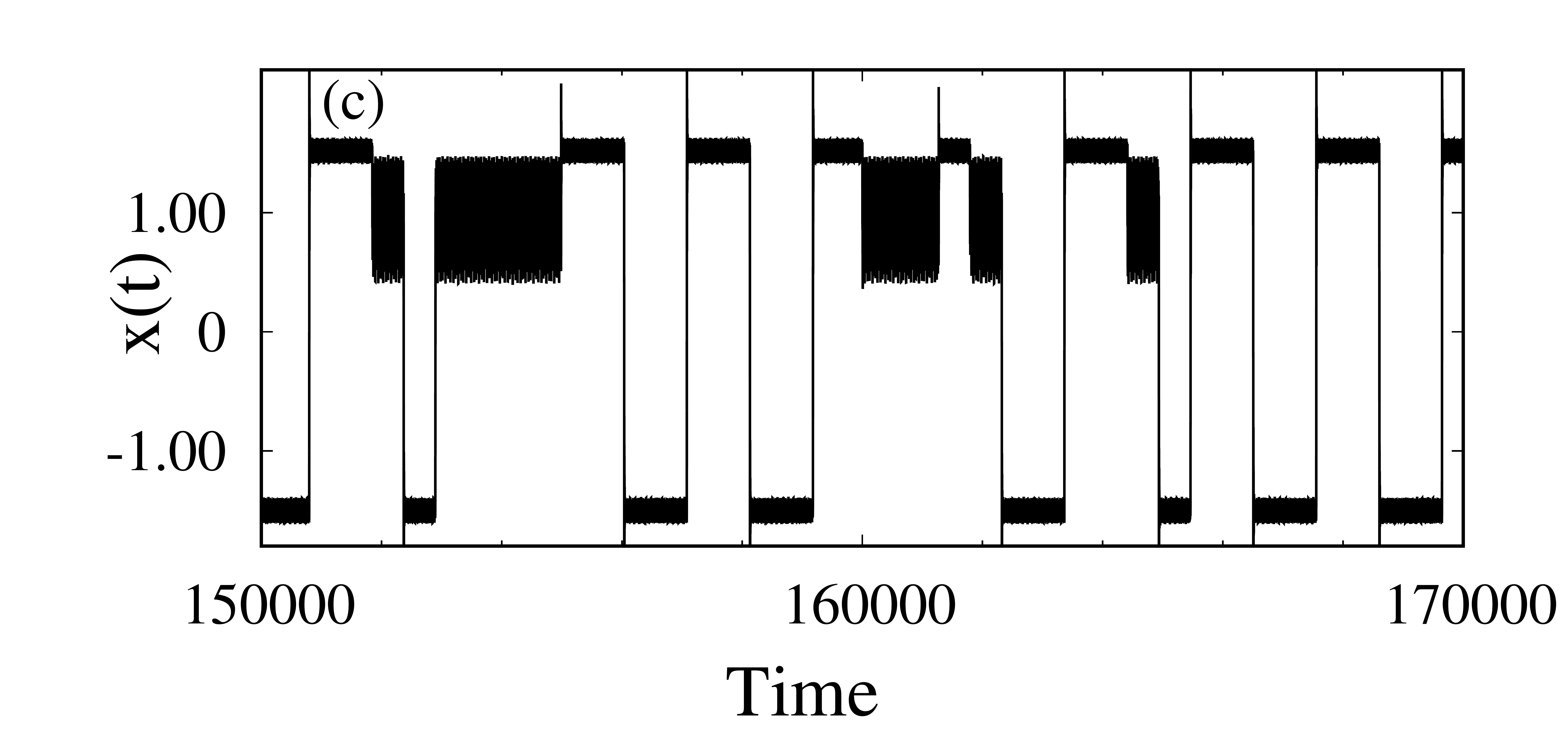} 
	\caption{Panel (a)shows two different streams of input signals of '3' level square waves. The input set for solid line shows $I=I_{1}+I_{2}$=-1.0 when the logic input is $'0'$ and if $I=I_{1}+I_{2}$=1.0 when the logic input is $'1'$, while dashed line shows  $I=I_{1}+I_{2}$=-2.0 when the logic input is $'0'$ and $I=I_{1}+I_{2}$=2.0 when the logic input is $'1'$. Panels (b) and (c) represent the dynamical response of the system for different inputs namely dashed and solid lines in (a), respectively, under quasiperiodic forcing for A =0.27 when one obtains the desired OR logic outputs for $\varepsilon=0.05$ (see Table \ref{Tab1}).}
	\label{fig9}
\end{figure} 

Next, the parameter space of the strength of external quasiperiodically forcing A and the amplitude of the square waves $\delta$ is scanned numerically in the range $A \in(0.0,0.4)$ and $\delta \in(0.0,1.0)$ to pinpoint different dynamical behaviours, and more specifically the occurrence of logical SNA, where the above discussed binary logic is valid. To start with, we demarcate the parameter space $(A,\delta)$ by numerically integrating Eq.\eqref{equ1} into quasiperiodic attractor, logical SNA, standard SNA and chaotic attractor by using Lyapunov exponents, phase sensitivity exponents and power spectral measures.

A two parameter numerical phase diagram is shown in Fig.\ref{fig8} for $A\in(0.0,0.4)$ and $\delta\in(0.0,1.0)$. The various dynamical behaviours indicated in the phase diagram and the interesting dynamical transitions are elucidated in the following. For low A and low $'\delta'$ values in the chosen range, the system exhibits quasiperiodic oscillations in one well, while for higher values of $\delta$ (and same low A values) it exhibits a torus behaviour encompassing both the wells. When the value of A exceeds a critical value, the quasiperiodic oscillations lose their smoothness and the attractor becomes a fractal torus/SNA. In specific ranges of parameter values of A, the system exhibits logical SNA and standard SNA for almost all $\delta$ values. However, for sufficiently large A and low values of $\delta$, the system is forced to behave chaotically as shown in Fig.\ref{fig8}. When considering the effect of $\delta$, it is observed that there are two types of transitions which are predominant here as A is increased. 
\begin{itemize}
	\item \textbf{Transition 1:} One well quasiperiodic oscillations $\rightarrow$ logical SNA $\rightarrow$ standard SNA $\rightarrow$ chaos.
	\item \textbf{Transition 2:} Two well quasiperiodic oscillations $\rightarrow$ logical SNA $\rightarrow$ standard SNA.
\end{itemize}

Now considering the logical SNA regions, we have already seen from Fig.\ref{fig6}, how a low amplitude $\delta$ is sufficient to realize the logic OR gate as well as AND gate (see the following section). However from Fig.\ref{fig9} it is observed that for different $\delta$ values of the input square wave, we can get the logical SNA region in a rather wide range of the parameter space. In digital electronics, the physical nature of the input and output signals/values of logical gates are expected to have the same ranges of values. To test the validity that the logic gates emerge in our system irrespective of the strength of the amplitude $\delta$ of the input signals (low/moderate), we vary the amplitudes of the input square waves $(I_{1}, I_{2})$ to 0.5 and 1.0 and we observe that the outputs of the system exhibit the same logical output values irrespective of the input values (Figs.\ref{fig9}). In fact we do find that the logic gates emerge almost in the entire region of $\delta \in [0.03,1.0]$. This confirms that our study is suitable for designing logic gates irrespective of the input values in the given logical SNA range as long as they are of low/moderate values.

\begin{figure}[]
	\centering	
	\includegraphics[width=0.9\linewidth]{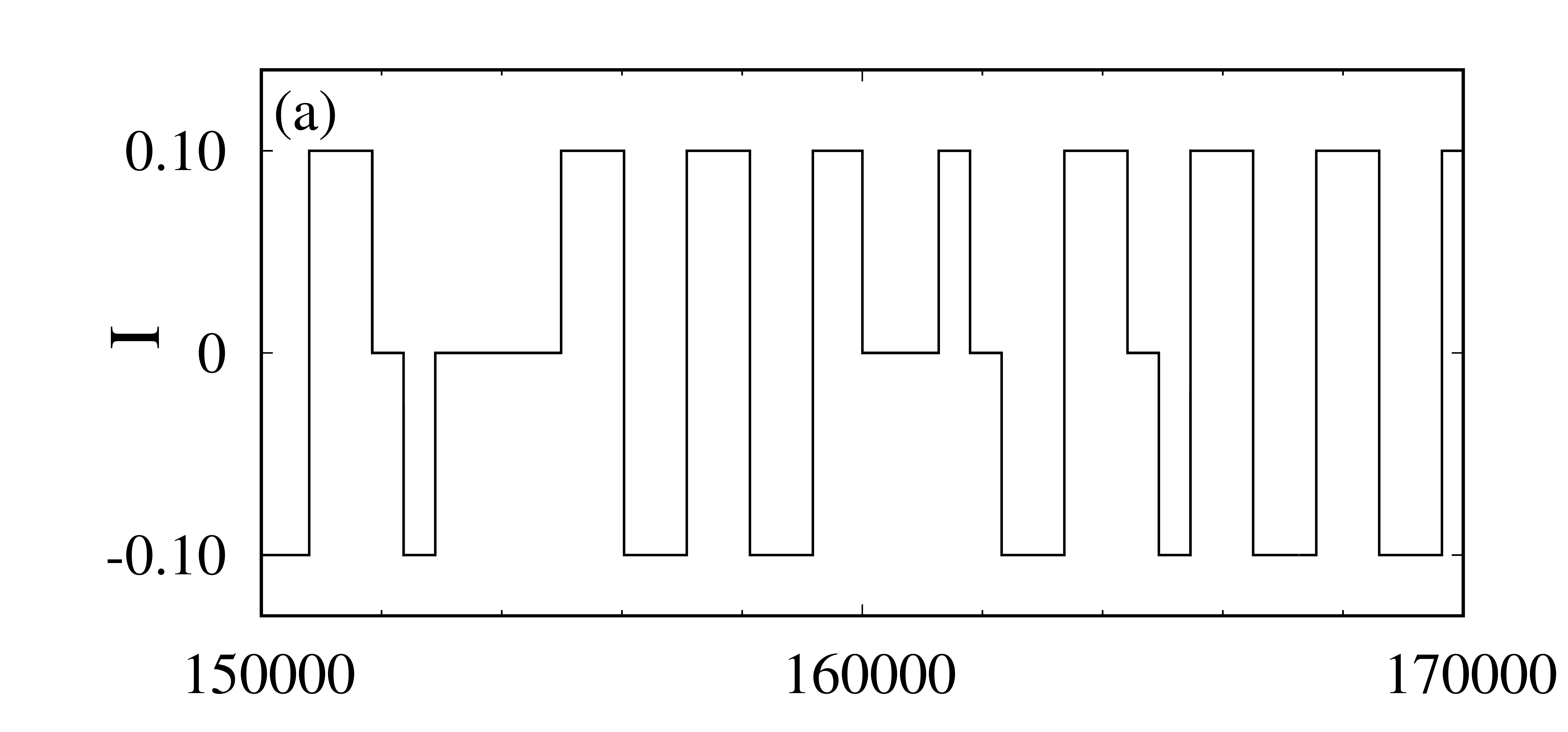} \\
	\includegraphics[width=0.9\linewidth]{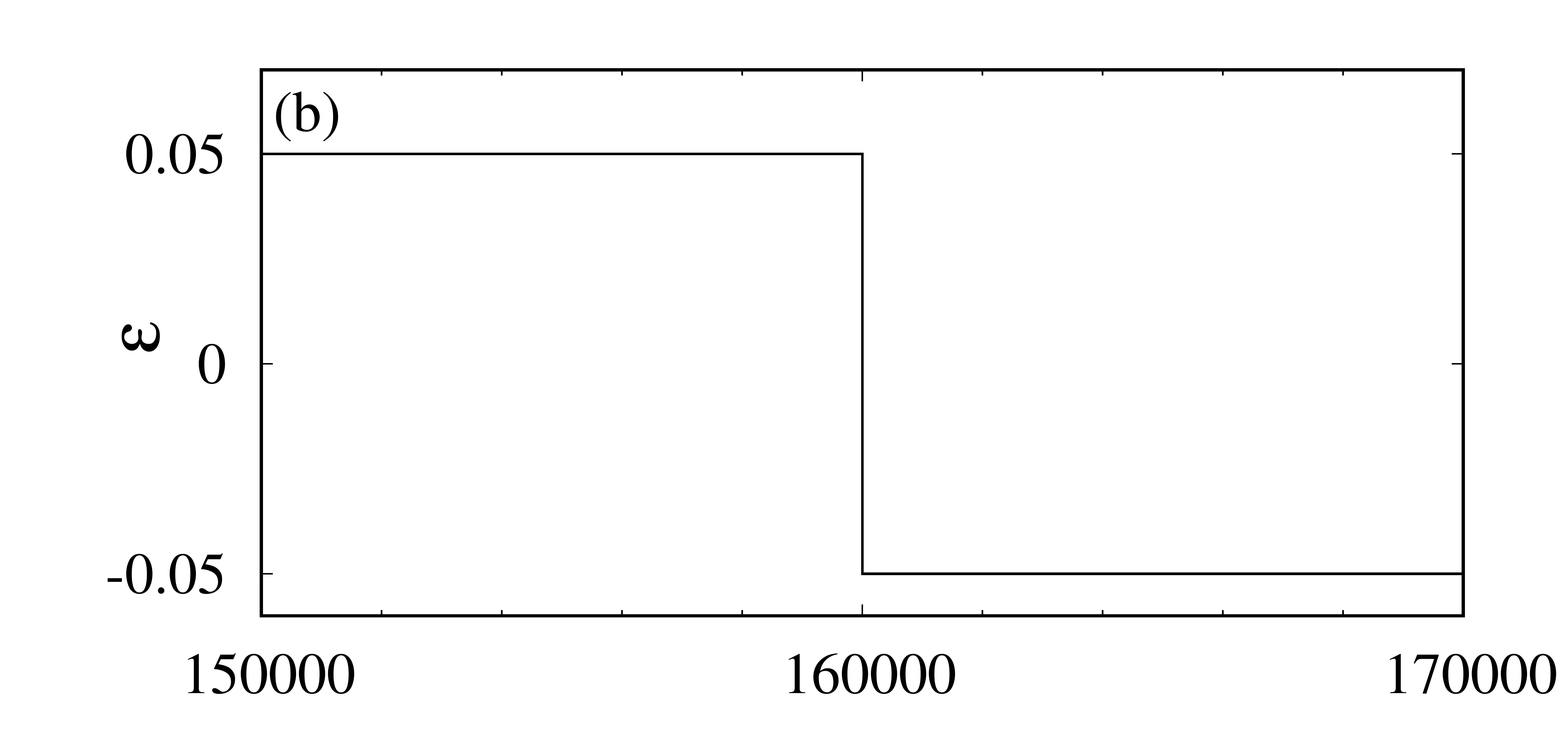} \\
	\includegraphics[width=0.9\linewidth]{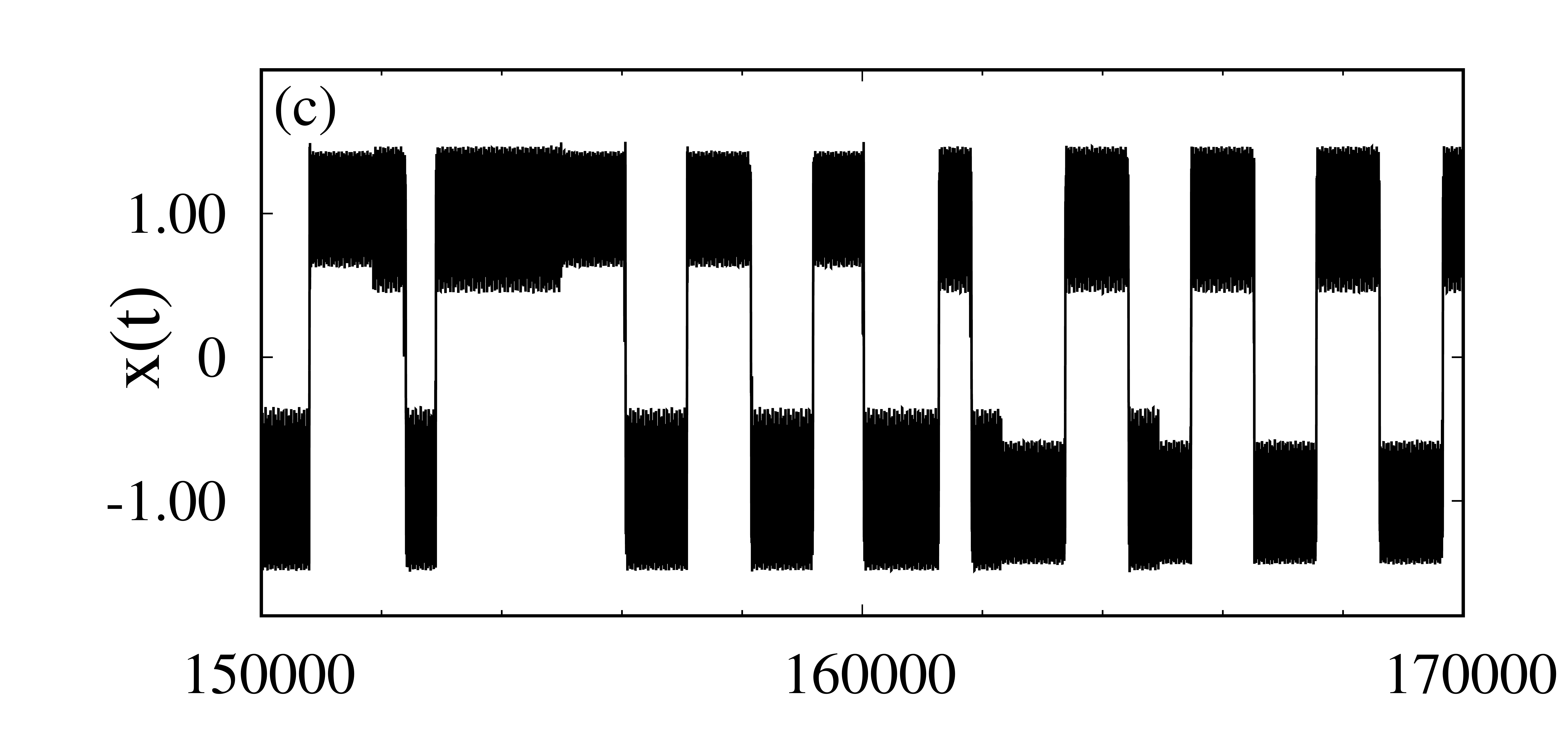} 
	\caption{From top to bottom panels: (a) shows a stream of '3' level square waves with -0.1 corresponding to the input set (0,0), 0 for (0,1)/(1,0) set and 0.1 for (1,1) input set. Panel (b) shows the asymmetric bias $\varepsilon=0.05$ leads to the desired OR logic and $\varepsilon=-0.05$ gives AND logic output. Panel (c) represent the dynamical response of the system under quasiperiodic forcing for A =0.27. Note that the quasiperiodic forcing is optimum when $A=0.27$ [panel(c)] where one obtains the desired OR/AND logic outputs for $\varepsilon=0.05/-0.05$ (see Table \ref{Tab1}).}
	\label{fig10}
\end{figure}
\begin{figure}
	\centering	
	\includegraphics[width=0.9\linewidth]{fig6a} \\
	\includegraphics[width=0.9\linewidth]{fig6b} \\
	\includegraphics[width=0.9\linewidth]{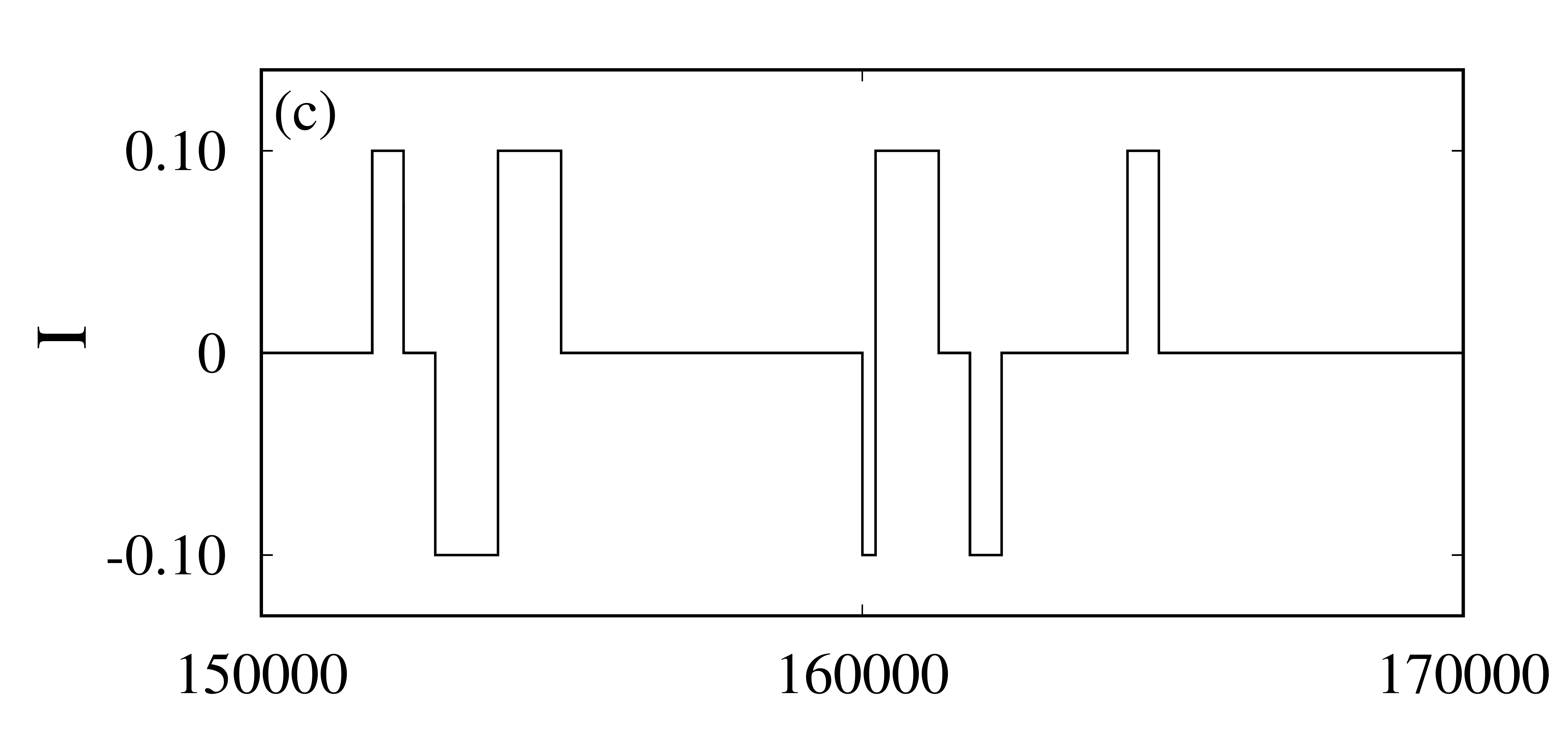} \\
	\includegraphics[width=0.9\linewidth]{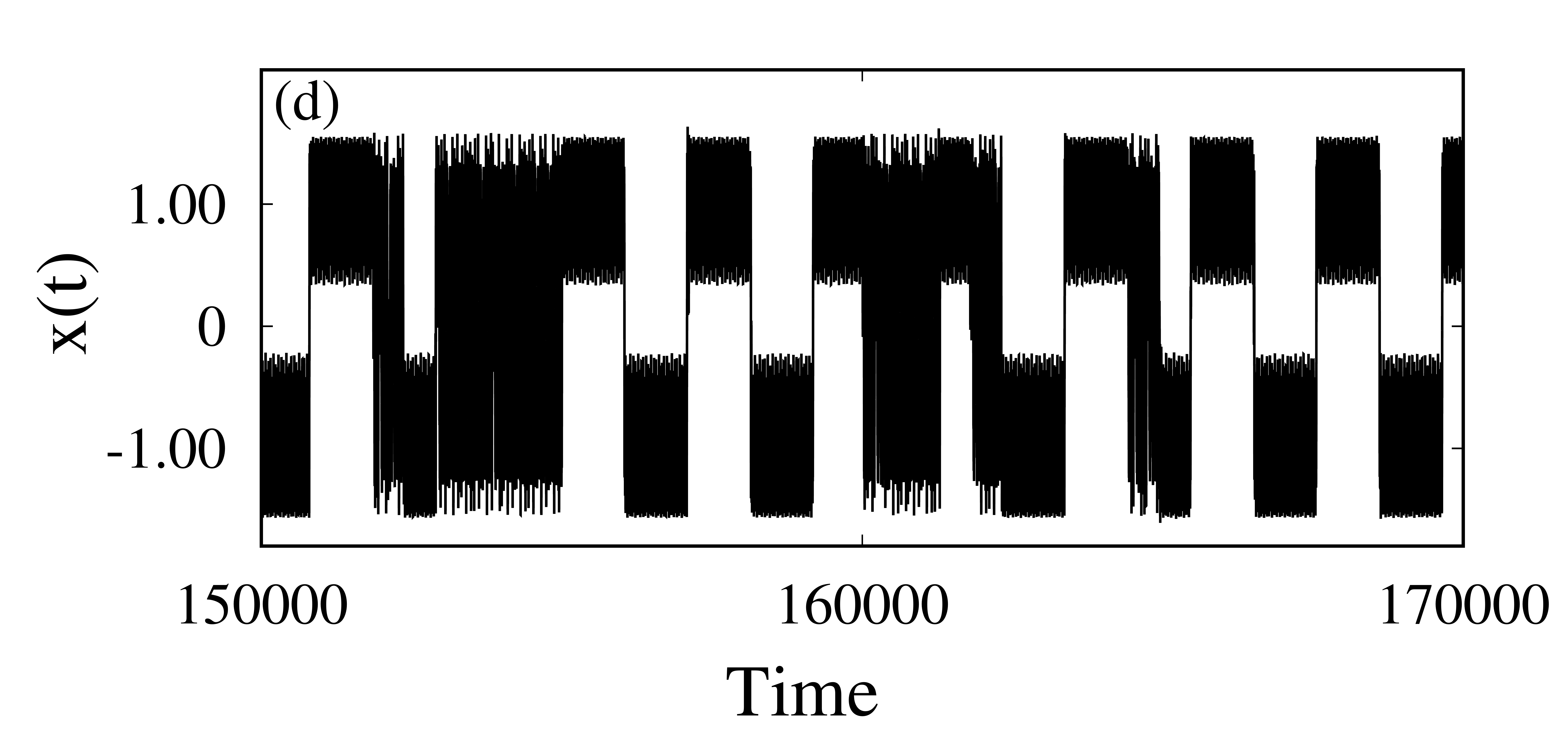}
	\caption{Panels (a)\&(b) show the stream of input signals $I_{1},I_{2}$. Panel (c) shows the '3' level square wave while panel (d) shows the desired  S-R latch when $\varepsilon=0.0$ with corresponding forcing A=0.32.}
	\label{fig11}
\end{figure}

\section{Implementation of other logic gates and effect of noise}
Finally in this section, we point out how the remaining logic gates, namely AND and NAND can be identified. Also we show how the system \eqref{equ1} can be used as a Set-Reset (SR) flip-flop. We further point out the reliability of obtaining logic gates in a quantitative way and the persistence of  gates even in the presence of noise.
\subsection*{A. AND and NAND gates}
We now study studied the effect of a different constant input bias $\varepsilon$ in (1). The results are displayed in Fig.\ref{fig10} . We observe that as the value of bias changes from $\varepsilon=0.05$ to $\varepsilon=-0.05$ the response of the system morphs from OR gate SNA to AND gate SNA logic behaviour as shown in Fig.\ref{fig10}(c). Here we notice that changing $'\varepsilon'$ causes an alternation in the symmetry of potential wells which leads to emulate different logical SNA responses. Similarly, we note that when the NOR gate bias is changed from $\varepsilon=0.05$ to $\varepsilon=-0.05$ the NAND gate logic emerges (the figures of which we do not specifically display here for brevity).

\begin{table}
	\begin{center}
		\caption{Truth Table of Set-Reset (SR) flip-flop}
		\vspace{0.2cm}
		\begin{tabular} {|c| c| c|}
			\hline
			Set($I_{1}$) & Reset ($I_{2}$) &Latch \\
			\hline
			0  & 0 & No change \\
			\hline
			0  & 1  & 0 \\
			\hline
			1  & 0  & 1 \\
			\hline
			1  & 1 & Restricted set  \\
			\hline
		\end{tabular}
		\label{Tab2}
	\end{center}
\end{table}

\subsection*{B. Set-Reset (SR) flip-flop}
Next to use this system as a Set-Reset (SR) flip-flop, we need to modify the encoding of input values. From the R-S flip-flop truth table given in Table \ref{Tab2}, it is very obvious that two states $(0,1)$ and $(1,0)$ yield different outputs. This may be accomplished by encoding in a different way so that the first input $I_1$ takes the value $-I$ when the logic is $0$ and $+I$ when the logic is $+1$. Similarly the second input $I_2$ takes the value $+I$ when the logic input is $0$ and $-I$ when the logic is $+1$. This will be implemented by applying NOT operation to the second input $I_2$. As a result for the 4-set of binary inputs $(I_1,I_2):(0,0),(0,1),(1,0),(1,1)$, the input signal $I$ takes the value $0,-1,1,$and $0$ respectively. Out of these four sets, $(1,1)$ set   is a restricted one. Hence  one has only a three-set input signal, thereby a three state level signal will be given to input (Fig.\ref{fig11}(c)). Logic response for output can be obtained as in the case of logic operations: $x>0$, the logic output is taken as $1$, $0$ for $x<0$. It is clearly evident from Fig.\ref{fig11}(d) that low/moderate quasiperiodic forcing consistently fields the Set-Reset latch input-output operations.
 
\subsection*{C. Quantification of reliability of obtaining logic gates}
\begin{figure}
	\centering
	\includegraphics[width=0.5\linewidth]{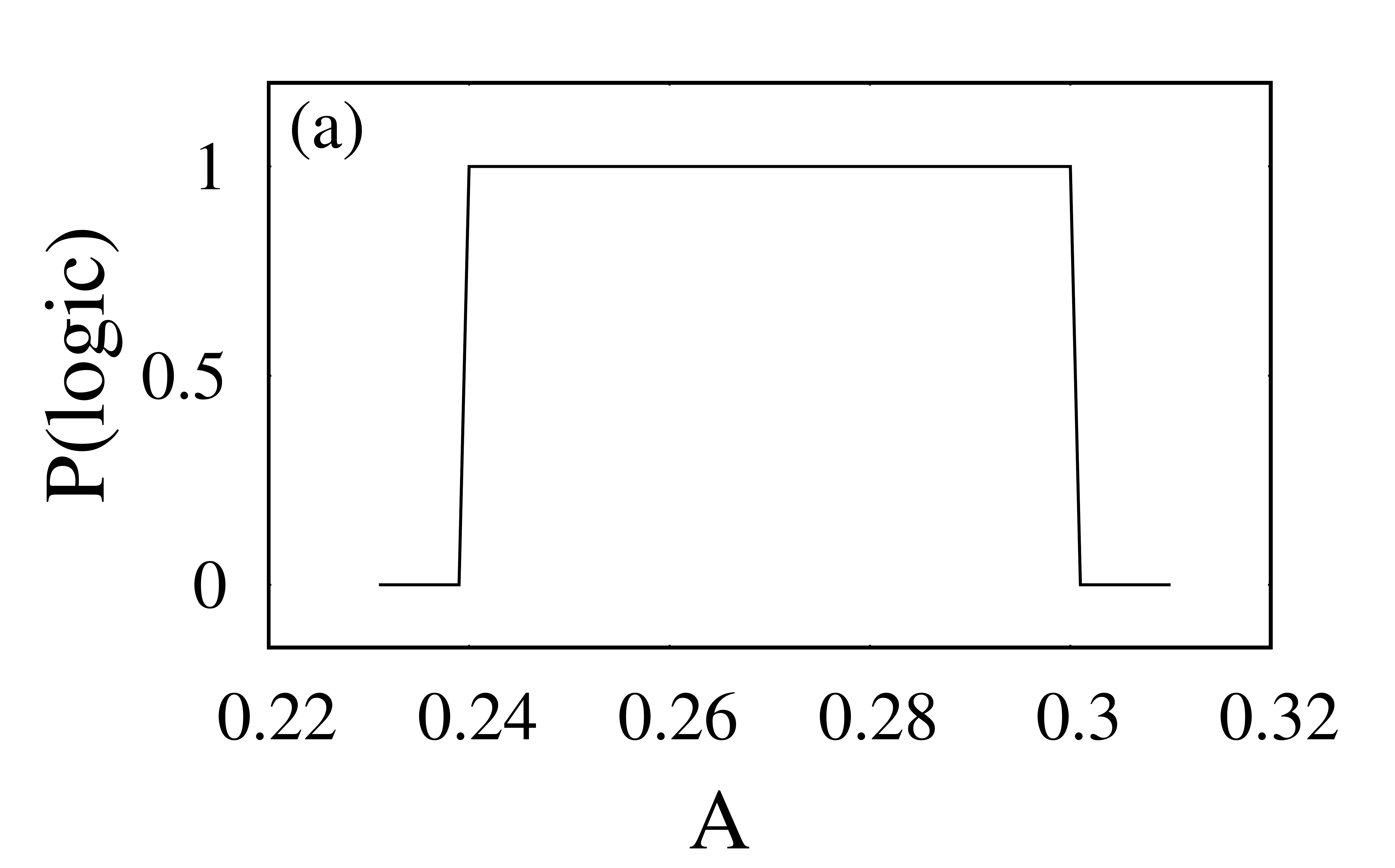}~
	\includegraphics[width=0.5\linewidth]{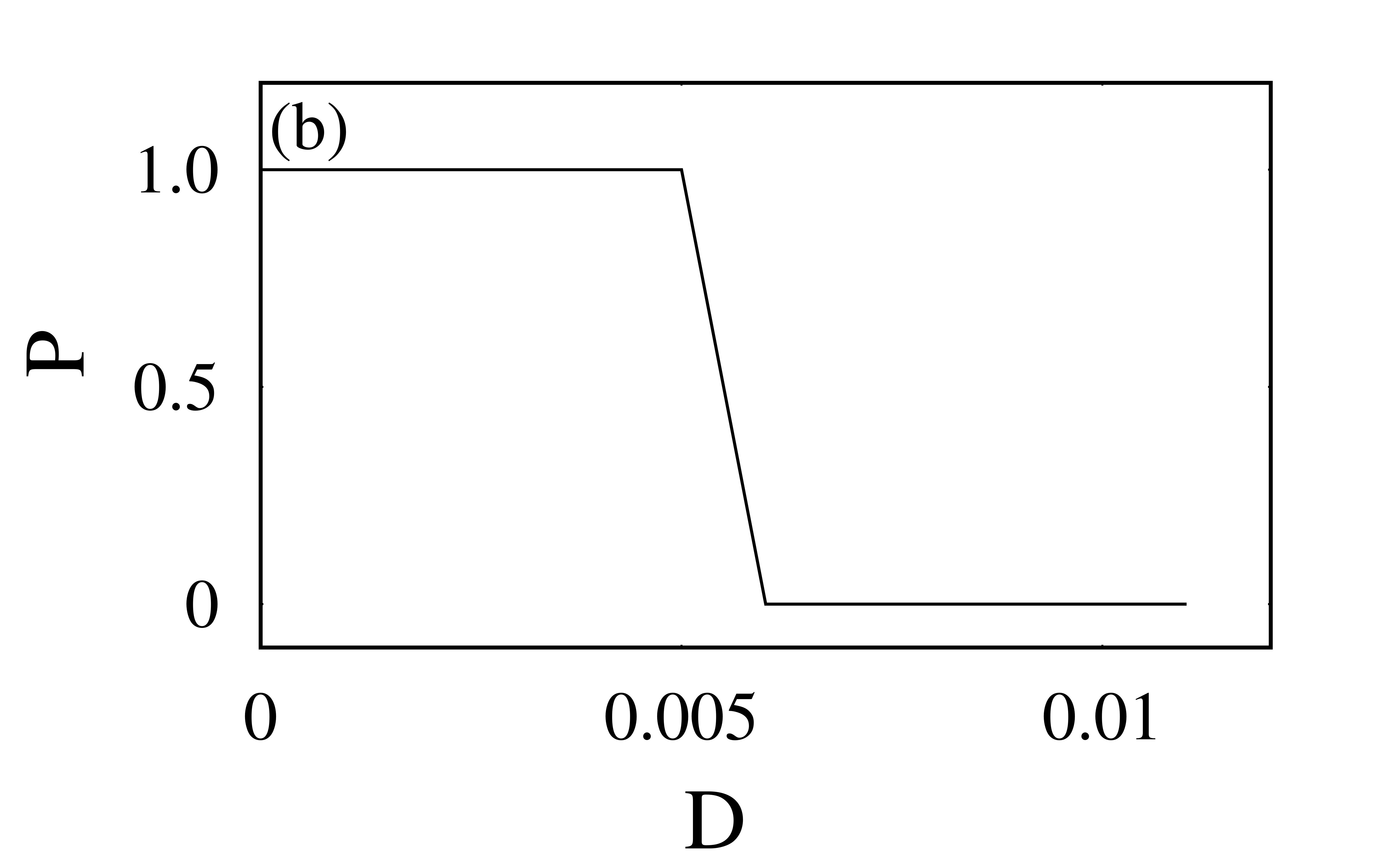}
	\caption{(a) Probability of obtaining the OR/AND operation for different values of quasiperiodic forcing $A$ with $\varepsilon=0.05/-0.05$, (b) Probability of obtaining OR/AND logic behaviour for different values of noise strength $'D'$ with fixed $A=0.27$.}
	\label{fig12}
\end{figure}

We can quantify the consistency/reliability of obtaining a given logic output by calculating the probability of obtaining the desired logic output for different sets of inputs, that is the ratio of the number of successful runs  to the total number of runs. Thus we define P(logic)to be $1$, when the logic operation is completely obtained for all given input sets, otherwise it is $0$. In the present case the system \eqref{equ1} was simulated by keeping the value of one input set constant over 1000 time steps and it is continued for a sequence of 500 such sets. It is evident from Fig.\ref{fig12}(a), we obtain a window for the quasiperiodic forcing for which our system consistently gives the desired logic responses as output. 

\subsection*{D. Effect of noise in the logic gates}
Finally, we consider the behaviour of system \eqref{equ1} in the presence of noise. For the sake of definiteness, we choose the noise intensity to be comparable to that of a weak internal noise which originates in electronic components that may model the system \eqref{equ1}. Such noise essentially originates in the analog components and is usually $\sim 1 \mu V$ \cite{luchinsky1998analogue, *khovanov2000effect, horowitz1980art}. It is observed that the behaviour of the largest Lyapunov exponent of the forcing amplitude $A$ in the presence of noise is found to be practically coincident with that of the noise free case. Hence, in the presence of noise, the logical SNA in the system retains its negative Lyapunov exponents and the fractal structure and that the behaviour of 'OR/AND' logic is possible for $D<0.005$ [see Fig.\ref{fig12}(b)]. Our study confirms that the logic behaviour remains when the noise strength is below $mV$ range. Hence the logic nature in our system persists even when the noise originates due to analog electronic components of the system.

\section{Conclusion}

In this paper, we have studied the response of quasiperiodically driven double-well Duffing oscillator to deterministic input signals. We have shown that if one uses two square waves in an aperiodic manner as input signals to the oscillator system, the response of the oscillator can produce logical SNA output controlled by the quasiperiodic forcing. Changing the threshold or biasing the oscillator changes the 'OR' logical SNA output to 'AND' logical SNA output (and similarly NOR to NAND) and SR flip-flop. We have also shown that the two distinct dynamical phenomena, namely SNA and computation, commonly thought of as arising under very different contexts in the study of nonlinear systems can actually be closely related. In the present work, we have shown that with a low/moderate quasiperiodic forcing, logic operations can be obtained in nonlinear dynamics subjected to two aperiodic square waves. The dynamical behaviour in the logic operation region is SNA. Consequently the dynamics is robust under weak noise. Therefore an efficient computational process can be designed.   

\section*{Acknowledgement} 

The work of MSA and ML forms part of a research project sponsored by DST-SERB under Grant No. EMR/2014/001076. ML also acknowledges the financial support under the DST-SERB Distinguished Fellowship program.

\end{document}